\documentclass[prd,preprint,tightenlines]{revtex4}

\usepackage{amsmath,amssymb,bm,color,graphicx,multirow,mfirstuc}
\usepackage[colorlinks]{hyperref} % not compatible with preview-latex ?
\hypersetup{
	citecolor=blue, 
	linkcolor=blue,
	urlcolor=blue
}

\usepackage{comment}
\includecomment{todo2}
\includecomment{test}

\usepackage[textsize=small]{todonotes}
\setuptodonotes{color=red!30}

\definecolor{ao(english)}{rgb}{0.0, 0.5, 0.0}  % dark green
\definecolor{bostonuniversityred}{rgb}{0.8, 0.0, 0.0}  % dark red
\definecolor{blue(pigment)}{rgb}{0.2, 0.2, 0.6}  % dark blue
\definecolor{brown(traditional)}{rgb}{0.59, 0.29, 0.0} % brown

\begin{document} 
\title{Consistent explanation for the cosmic-ray positron excess in $p$-wave Sommerfeld-enhanced dark matter annihilation}

\author{Yu-Chen Ding$^a$}
\author{Yu-Lin Ku$^a$}
%\author{Yu-Lin Ku$^a$}
\author{Chun-Cheng Wei$^a$}
\author{Yu-Feng Zhou$^{a,b}$}%\email{yfzhou@itp.ac.cn}

\affiliation{$^a$CAS key laboratory of theoretical Physics, Institute of Theoretical Physics, Chinese Academy of Sciences, Beijing 100190, China; 
School of  Physics, University of Chinese Academy of Sciences, Beijing 100049, China;
$^b$School of Fundamental Physics and Mathematical Sciences, Hangzhou Institute for Advanced Study, UCAS, Hangzhou 310024, China; 
International Centre for Theoretical Physics Asia-Pacific, Beijing/Hangzhou, China.}
\date{\today}

\begin{abstract}
Sommerfeld-enhanced  dark matter (DM) annihilation through $s$-wave has been widely considered as a consistent explanation for both the observed cosmic-ray (CR) positron excess and the DM thermal relic density. 
%as in this mechanism the DM annihilation cross section is enhanced towards lower DM  velocities.
%
However, as the  $s$-wave Sommerfeld-enhanced annihilation cross section  increases monotonically  with decreasing DM velocity,  severe constraints appear  from the data of gamma rays from dwarf spheroidal satellite galaxies (dSphs) and the cosmic microwave background (CMB), as the relevant typical  DM velocities  are even lower than that in the Galactic halo.
In this work, we consider Sommerfeld-enhanced $p$-wave DM annihilation where
the DM annihilation cross section can be enhanced at certain velocities but eventually be highly suppressed when the DM velocities become extremely low.
We calculate the velocity-dependent astrophysics factors ($J$-factors) for the Sommerfeld-enhanced  $p$-wave DM annihilation  for fifteen nearby dSphs.
%
%TODO 4.a
Taking the channel  of DM annihilating into $4\mu$ through two light mediators as an example,  
we show that there are parameter regions where this mechanism can   account for  the CR positron excess and the  DM relic density, while being compatible with the constraints from dSphs gamma rays measured by  Fermi-LAT and that from the CMB measured by PLANCK. 
\end{abstract}
\date{\today}
\maketitle

\section{Introduction}
Numerous astrophysical observations indicate that non-baryonic cold dark matter (DM) constitutes $\sim 84\%$ of the total matter of the Universe~\cite{Aghanim:2018eyx}. However, the particle nature of DM remains largely unknown. Weakly-interacting massive particles (WIMPs) are one of the popular DM candidates which can naturally  obtain the observed DM  abundance through self-annihilation into the standard model (SM) final states during  the process of thermal freeze-out.  
If DM particles in the Galactic halo can annihilate into SM stable final states, they can make extra contributions to the fluxes of cosmic-ray (CR) particles which can be probed by the current DM indirect detection experiments.

CR antiparticles, such as CR positrons and antiprotons, are considered to be relatively rare as they are CR secondaries produced dominantly from the collisions between  primary CRs and the interstellar gas. Thus they are expected to be sensitive to extra contributions, and are important probes of DM interactions.
In recent years, a number of experiments such as PAMELA~\cite{Adriani:2013uda}, Fermi-LAT~\cite{FermiLAT:2011ab} and AMS-02~\cite{Aguilar:2019owu} have reported an excess of CR positron flux starting at positron energy above $\sim10~ \mathrm{GeV}$ and peaking at  $\sim300$~GeV, which strongly suggests the existence of nearby extra positron sources.
The excess may have astrophysical origins, such as nearby pulsar wind nebulae (see e.g.~\cite{Hooper:2008kg, Yuksel:2008rf, Profumo:2008ms,Hooper:2017gtd}), and supernovae remnants (e.g.~\cite{Blasi:2009hv, Hu:2009bc, Fujita:2009wk}), etc.. DM annihilation or decay in the Galactic halo can also be a possible explanation~\cite{Bergstrom:2008gr,Cirelli:2008pk,Bergstrom:2009fa,Lin:2014vja,Jin:2013nta,Jin:2014ica}.
In the scenario of DM annihilation, the favored DM typical particle mass is $m_\chi\sim \mathcal{O}(0.5-1)$~TeV and velocity-weighted annihilation cross section $\langle \sigma_{\rm{ann}} v_{\text{rel}} \rangle \sim \mathcal{O}(10^{-24}) ~\text{cm}^3\ \text{s}^{-1}$, which depends on annihilation channels~\cite{Cholis:2008hb,Cirelli:2008pk,Bergstrom:2009fa,Cirelli:2009dv}.
The favored DM parameter space is, however, strongly constrained by other observations. 
For instance, DM annihilation directly into quark and gauge boson  pairs ($\bar q q$, $W^+W^-$ and $Z^0Z^0$) is ruled out  by the lack of corresponding excesses in the CR antiproton flux~\cite{Aguilar:2016kjl}.
For leptonic channels, DM particles dominantly annihilating into $2\tau/4\tau$ leads to a good fit to the AMS-02 data. However, this channel produces significant amount of gamma rays, which is ruled out by Fermi-LAT  observations on gamma rays from  dwarf spheroidal satellites galaxies (dSphs)~\cite{Ackermann:2015zua}.
The $\mu^+\mu^- \ (2\mu)$ and $\mu^+\mu^-\mu^+\mu^- \ (4\mu)$ channels produce much less gamma rays and thus only have mild tensions with the Fermi-LAT gamma-ray data. Compared with the $2\mu$ channel, the  $4\mu$ channel  predicts a broader positron excess, which is in a better agreement with the AMS-02 data.

If the DM annihilation cross section remains constant (velocity independent) during different epochs of the Universe, there exist other severe constraints. 
For instance, 
for  WIMP DM candidates, very strong constraint arises from the DM relic density.
The same DM  annihilation cross section at the epoch of thermal freeze-out should be responsible for its relic density, which sets a typical scale of $\langle \sigma_{\rm{ann}} v_{\text{rel}} \rangle \sim 3\times 10^{-26}~\text{cm}^3\cdot\text{s}^{-1}$,
which is about two orders of magnitude lower than that   required to explain the observed  CR positron excess, which is often referred to as the ``boost factor'' problem for DM explanation~\cite{Cholis:2008hb,Cirelli:2008pk}. It is very unlikely that the local DM density clumps can provide such a large boost factor~\cite{Diemand:2008in,Springel:2008by}.
During the epoch of recombination, the charged particles produced by DM annihilation  can modify the cosmic microwave background (CMB)~\cite{Chen:2003gz,Padmanabhan:2005es,Slatyer:2015kla,Slatyer:2015jla,Galli:2009zc}. The precise measurements of CMB temperature and polarization anisotropy performed by Planck have set upper limits on DM annihilation cross section~\cite{Aghanim:2018eyx}.  For some DM annihilation channels, the constraints from CMB are more stringent than that derived from  dSphs gamma-ray observations.

Since the typical velocity of DM particles varies significantly in different epochs and regions of the Universe, a possible solution to reconcile these  apparent tensions is to introduce velocity- or temperature-dependent DM annihilation cross sections.
An extensively studied mechanism is the Sommerfeld enhancement~\cite{https://doi.org/10.1002/andp.19314030302,Hisano:2003ec,Hisano:2002fk,Cirelli:2007xd,ArkaniHamed:2008qn,Pospelov:2008jd,MarchRussell:2008tu,Iengo:2009ni,Cassel:2009wt}.
The Sommerfeld enhancement of annihilation cross section occurs when the annihilating particles self-interact through a long-range attractive potential at low velocities. In this scenario, the short-distance DM annihilation cross-section can be greatly enhanced due to the distortion of the wave functions of annihilating particles at origin.  The attractive potential may originate from multiple-exchange of light force-carrier particles between the annihilating DM particles.
The $s$-wave Sommerfeld enhancement has been considered to reconcile the apparent tension between CR positron excess and the thermal relic density in WIMP scenarios, as the DM typical relative velocities $v_{\text{rel}}$ in the Galactic halo and at the freeze-out are $\sim 10^{-3}c$ and $\sim 10^{-1}c$, respectively~\cite{Hisano:2003ec,Hisano:2006nn,Cirelli:2007xd,ArkaniHamed:2008qn,Lattanzi:2008qa,Liu:2013vha}, where $c$ is the speed of light.
For simple models with a vector mediator, the allowed parameter space is quite limited, as the DM annihilation cross section also receives contributions from DM annihilating into light mediators~\cite{Feng:2010zp}. But for more complicated models, such as the exciting DM model there still exists large parameter space to explain the data~\cite{Finkbeiner:2010sm}.
Note, however, that the  $s$-wave Sommerfeld-enhanced annihilation cross section  increases monotonically  with decreasing DM velocity, which makes it even more difficult to  reconcile other constraints  such as that from the data of  dSphs gamma rays and CMB, as the related typical velocity is even lower $\sim10^{-4}c$ and $\sim10^{-8}c$, respectively~\cite{Zavala:2009mi,Hannestad:2010zt,Hisano:2011dc,Kamionkowski:2008gj,Bovy:2009zs,Buckley:2009in,Feng:2009hw,Cholis:2010px,Lattanzi:2008qa,Robertson:2009bh,Cirelli:2010nh,Abazajian:2011ak}.

In this work, we consider DM particles annihilating through $p$-wave with Sommerfeld enhancement. 
The $p$-wave Sommerfeld-enhanced cross section no longer  increases monotonically  towards lower DM velocity, since the cross section before the enhancement  has significant velocity dependence $\sigma_{\rm{ann}} v_{\text{rel}}\sim v_\text{rel}^2$. 
After including the effect of Sommerfeld enhancement, the DM total annihilation cross section can be enhanced at some velocities, but eventually be suppressed towards the limit $v\to 0$. Thus it opens the  possibility of reconciling the above mentioned tensions.
In this work, we consider a typical DM annihilation channel of $\chi\chi\to 2\phi\to4\mu$, where $\phi$ is the mediator particle and also the force-carrier. 
We calculate the velocity-dependent astrophysics factors ($J$-factors) for the Sommerfeld-enhanced  $p$-wave DM annihilation through a light mediator $\phi$ for 15 nearby dSphs.
A combined  analysis on the $p$-wave Sommerfeld enhanced DM annihilation in CR positron excess, DM thermal relic density, gamma-ray constraints from dSphs and CMB constraints is performed.  We  find that a consistent explanation for all these observables is possible in this scenario.

%%
%Note that the Gamma-ray observations from other regions of the sky can also place stringent limits on the DM annihilation, such as that from the Galactic center (GC)  observed by H.E.S.S.~\cite{Abdallah:2016ygi}, and the isotropic gamma ray background (IGRB) measured by Fermi-LAT~\cite{Ackermann:2015tah}. As these constraints have additional uncertainties  due to their strong dependences on the DM density profile on galactic and cosmic scales~\cite{HESS:2015cda,Abdo:2010dk}, we do not consider these constraints in the present work.

%TODO 2  
Note that the gamma-ray observations from other regions of the sky can also place stringent constraints on the DM annihilation, such as that from the Galactic center (GC) observed by H.E.S.S.~\cite{Abdallah:2016ygi}, and the isotropic gamma ray background (IGRB) measured by Fermi-LAT~\cite{Ackermann:2015tah}. However, these constraints  involve additional uncertainties.
As shown in Ref.~\cite{HESS:2015cda}, changing the DM density profile in the Galactic halo from a cusped NFW or Einasto profile to a cored NFW or Einasto profile for a core radius of 500~pc would result in the  limits on the DM annihilation cross section to be around two orders of magnitude weaker. A larger core radius would lead to weaker H.E.S.S. constraints.
%
%% IGRB
A significant  uncertainty in constraining DM properties from IGRB arises 
from the  modeling of the expected DM annihilation luminosity, 
which depends on the history of DM formation at very large scales.
Base on the Millennium II (MS-II) N-body simulation of cosmic structure formation~\cite{Boylan-Kolchin:2009alo}, 
the constraints on DM annihilation from IGRB signals in several scenarios of  the expected DM annihilation luminosity was discussed in Ref.~\cite{Fermi-LAT:2010qeq}.
The results showed that the limits  in the scenario with only the DM signals from halos/subhalos resolved in MS-II (with mass $\gtrsim 10^{6}M_\odot$, where $M_\odot$ is the mass of the Sun) considered are about three orders of magnitude weaker than that in the scenario where a extrapolation of the contribution from halos and subhalos down to $\sim 10^{-6}M_\odot$ was adopted.
Note  a very large N-body simulation with a high-resolution particle mass of $\sim 10^{-11}M_\odot$ by using a multi-zoom technique was recently completed~\cite{Wang:2019ftp}. It is possible that with   higher resolution simulations, the uncertainties from the modeling of the expected DM annihilation luminosity can be reduced.
Due to the still large uncertainties in the DM distribution at galactic and cosmological scales, we do not consider these constraints in the present work. 

This paper is organized as follows.
In Sec.~\ref{sec:ams}, we perform an updated  analysis on the AMS-02 CR positron data in DM annihilation scenario, and derive the favored  DM masses and annihilation cross sections.
In Sec.~\ref{sec:se&re}, we numerically calculate the $p$-wave Sommerfeld enhancement factors and compare them with the approximate analytic results in the literature. 
In Sec.~\ref{sec:relic-density}, the constraints from DM thermal relic density is discussed.
In Sec.~\ref{sec:Fermi}, we calculate upper limits on DM annihilation cross section from Fermi-LAT  gamma-ray data of 15 nearby dSphs.
In Sec.~\ref{sec:CMB}, we discuss the constraints from CMB measured  by Planck.
In Sec.~\ref{sec:combine}, we perform a numerical scan and identify the parameter regions which can explain all the relevant observations.
We summarize the conclusions of this work in Sec.~\ref{sec:conclusion}.

\section{AMS-02 positron excess and DM annihilation interpretations}\label{sec:ams}

%CR positrons are believed to be secondary particles produced from the scattering
%between primary CRs and the interstellar gas. CR secondary particles are relatively rare compared with the primary CRs, which makes them sensitive to extra contributions such as that from DM interactions.
%%
In recent years, an excess of CR positron flux at energy above $\sim10\ \text{GeV}$ has been observed by a number of experiments including PAMELA~\cite{Adriani:2013uda}, Fermi-LAT~\cite{FermiLAT:2011ab} and AMS-02~\cite{Aguilar:2019owu}, which can be possibly related to DM annihilation~\cite{Bergstrom:2008gr,Cirelli:2008pk, Yin:2008bs,Bergstrom:2009fa,Lin:2014vja,Jin:2013nta,Jin:2014ica}.
In this section, we use the CR positron flux data measured by AMS-02 to place
constraints on the DM annihilation cross sections. 
The propagation of CR particles in the Galactic halo can be modeled by a diffusion process within a diffusion zone which is a cylinder with radius $R_h=20\ \text{kpc}$ and half-height $Z_h=1\sim 10\ \text{kpc}$.
The diffusion equation of CR charged particles is given by~\cite{Ginzburg:1990sk,Strong:2007nh}
\begin{equation}
\frac{\partial \psi}{\partial t} = \nabla \left( D_{x x} \nabla \psi - \mathbf{V}_{c} \psi \right) + \frac{\partial}{\partial p} p^{2} D_{p p} \frac{\partial}{\partial p} \frac{1}{p^{2}} \psi - \frac{\partial}{\partial p} \left[\dot{p} \psi - \frac{p}{3} \left( \nabla \cdot \mathbf{V}_{c} \right) \psi\right] - \frac{1}{\tau_{f}} \psi -\frac{1}{\tau_{r}} \psi + q(\mathbf{r}, p)\ ,
\label{eq:dr}
\end{equation}
where $\psi(\mathbf{r}, p, t)$ is the CR particle's number density per unit momentum.
The spatial diffusion coefficient $D_{xx}$ can be written as $D_{xx} = \beta D_0 (R/R_0)^\delta$, where $R=p/(Ze)$ is the rigidity of CR particles with electric charge $Ze$,
$\beta=v/c$ is the velocity of CR particles, $D_0$ is a normalization constant, $\delta$ is the spectral power index, and $R_0$ is a reference rigidity.
$\mathbf{V}_c$ is the convection velocity.
$D_{pp} = 4 p^2 V_a^2 / (3 D_{xx} \delta (4-\delta^2)(4-\delta))$ is the diffusion coefficient in momentum space, where $V_a$ is the Alfvèn velocity which characterizes the propagation of weak disturbances in a magnetic field.
$\dot{p} \equiv dp/dt$ is the momentum loss rate of CR particles in propagation. 
$\tau_f$ and $\tau_r$ are the time scales of particle fragmentation and radioactive decay respectively.
The source term of primary CR particles can be written as $Q(\mathbf{r}, p, t)=f(\mathbf{r}, t)q(p)$, where $f(\mathbf{r}, t)$ is the spatial distribution taken from~\cite{1996A&AS..120C.437C}, and $q(p)$ is the momentum distribution.
The spectrum of momentum distribution is assumed to be a broken power law in rigidity $R$, $q(p)\propto (R/R_s)^{\gamma_p}$ with the spectral index $\gamma_p=\gamma_{p1}(\gamma_{p2})$ for the nucleus rigidity $R$ below (above) a reference rigidity $R_s$.
The boundary conditions are that the number density of CR particles vanishes at $r = R_{h}$ and $z = \pm Z_h$.
A steady-state solution can be obtained by setting $\partial \psi/\partial t=0$.
We use the public code \texttt{GALPROP v54}~\cite{Strong:1998pw,Moskalenko:2001ya,Strong:2001fu, Moskalenko:2002yx,Ptuskin:2005ax} to numerically solve this propagation equation.
We consider a number of propagation models listed in  Tab.~\ref{tab:propagation1}. These models are obtained from a global fit to the AMS-02 proton and B/C data using the \texttt{GALPROP} code, which represent the typically minimal ("MIN"), median ("MED"), and maximal ("MAX") CR fluxes in re-acceleration (DR) propagation model~\cite{Jin:2014ica}. We shall focus on  the ``MED" propagation model as a benchmark model.
\begin{table}[!h]
	\begin{tabular}{c c c c c c c c c}
		\hline \hline
		Model & $R_h\ (\rm{kpc})$ & $Z_h\ (\rm{kpc})$ & $D_0\ (\times 10^{28}\ \rm{cm^2\ s^{-1}})$ & $R_0\ (\rm{GV})$ & $\delta$ &$V_a\ (\rm{km\ s^{-1}})$ & $R_s\ (\rm{GV})$ & $\gamma_{p1}/\gamma_{p2}$ \\
		\hline
		MIN & 20 & 1.8 & 3.53 & 4.0 & 0.3 & 42.7 & 10.0	& 1.75/2.44 \\
		MED & 20 & 3.2 & 6.50 & 4.0 & 0.29 & 44.8 & 10.0 & 1.79/2.45 \\ 
		MAX & 20 & 6.0 & 10.6 & 4.0 & 0.29 & 43.4 & 10.0 & 1.81/2.46 \\
		\hline \hline
	\end{tabular}
	\caption{Parameters in the ``MIN'', ``MED'' and ``MAX'' propagation models derived from Ref.~\cite{Jin:2014ica}.	}
	\label{tab:propagation1}
\end{table}

The secondary CR positrons are produced from the interaction of primary CR particles with the interstellar gas. The corresponding source term is given by 
\begin{equation}
Q_{\mathrm{sec}}(p)=\sum_{i=\mathrm{H}, \mathrm{He}} n_{i} \sum_{j} \int c \beta_{j} n_{j}\left(p^{\prime}\right) \frac{d \sigma_{i j}\left(p, p^{\prime}\right)}{d p} d p^{\prime}\ ,
\end{equation}
where $n_i$ is the number density of the interstellar gas, $n_j(p^\prime)$ is the number density of CR particles, $d\sigma_{ij}(p,p^\prime)/dp$ is the differential cross section for the production of CR positrons.

The source term of the primary CR positrons produced by DM self-annihilation in the Galactic halo can be written as
\begin{equation}
q_{e^+}(\mathbf{r}, p) = \frac{\rho_\chi^2(\mathbf{r})\langle \sigma_{\text{ann}} v_{\text{rel}} \rangle}{2 m_{\chi}^{2}} \frac{d N_{e^+}}{d p}\ ,
\label{eq:positron.src.2}
\end{equation}
where $\rho_\chi$ is the DM energy density, $m_\chi$ is the DM mass,
$\langle \sigma_{\text{ann}} v_{\text{rel}} \rangle$ is the velocity-weighted DM annihilation cross section, and $d N_{e^+}/d p$ is the momentum distribution of positrons.
The spatial distribution of DM energy density $\rho_\chi$ in the Galactic halo is taken from the Einasto profile~\cite{Einasto:2009zd}
\begin{equation}
\rho_\chi(r) = \rho_\odot \exp \left[ - \left( \frac { 2 } { \alpha_E } \right) \left(\frac { r^{\alpha_E} - r_\odot^{\alpha_E}} { r_s^{\alpha_E} }\right)\right]\ ,
\label{eq:einasto-profile}
\end{equation}
with
$\rho_\odot = 0.43\ \text{GeV}\ \text{cm}^{-3}$, $\alpha_E = 0.17$,
$r_s = 20\ \text{kpc}$, and $r_\odot = 8.5\ \text{kpc}$~\cite{Salucci:2010qr}.

\begin{figure}[t]
\includegraphics[width=0.49\textwidth]{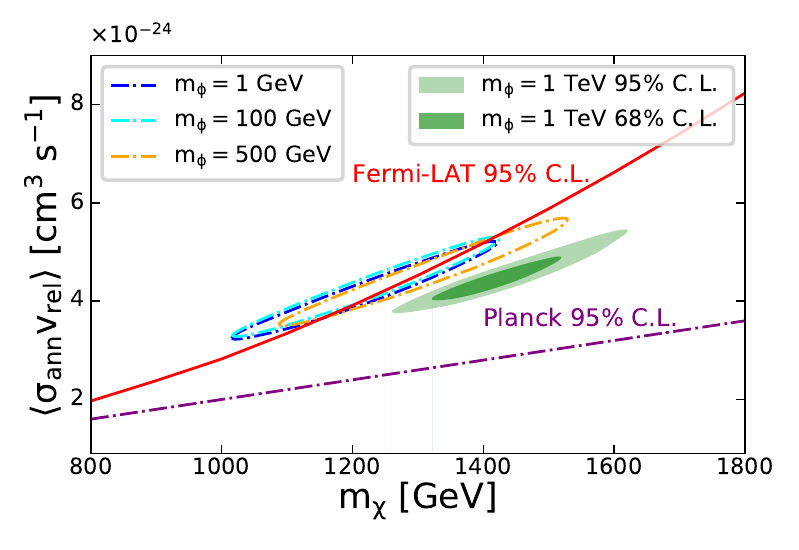}
\includegraphics[width=0.49\textwidth]{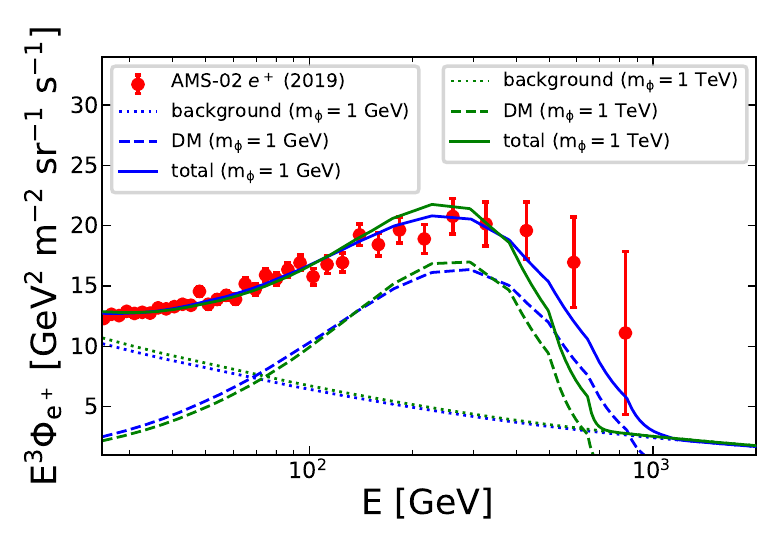}
\caption{
Left) Regions of DM mass and annihilation cross section favored by the AMS-02 positron flux data~\cite{Aguilar:2019owu} for the annihilation channel of $\chi\chi\to 2\phi \to 4\mu$.
The green contours  correspond to the heavy mediator case with $m_\phi = 1\ \rm{TeV}$, favored by AMS-02 positron data at 95\%~C.L. (outer) and 68\%~C.L. (inner). The blue, cyan and orange lines are for the cases with  mediator masses $m_\phi = 1$, 100, and 500~GeV, respectively.
The red and purple curves stand for upper limits at $95\%$ C.L. derived from the  Fermi-LAT data on dSphs gamma rays~\cite{Ackermann:2015zua} and the Planck  measurements of CMB~\cite{Aghanim:2018eyx}
%TODO 4.b
in the assumption of constant (velocity-independent) DM annihilation cross section.
See text for detailed explanation.
Right) Best-fit CR positron fluxes (solid lines) for two different mediator masses.
The dash and dot curves represent the contributions from DM annihilation and secondary positron backgrounds, respectively. The blue and green curves correspond to mediator masses of $m_\phi$=1~GeV and 1~TeV, respectively. }
\label{fig:ams-pos}
\end{figure}

We focus on the DM annihilation process of DM particles $\chi$ to $4\mu$ final state through a mediator $\phi$ with mass $m_\phi$, namely,  $2\chi\to 2\phi \to 4\mu$.
%

%TODO 1.2
	The scalar mediator can couple dominantly to leptons through mixing with leptophillic Higgs doublets in some variants  of the two-Higgs-doublet models~\cite{Batell:2016ove, DAmbrosio:2002vsn}. 
	If the mediator mass is smaller enough, the decay channel of $\tau^+\tau^-$ will be kinematically forbidden, and the $\mu^+\mu^-$ channel can be dominant.
	There are also  flavor specific scenarios which allow the mediator to couple dominantly to muons in a technically natural way~\cite{Batell:2017kty, Marsicano:2018vin}. 
The Monte-Carlo event generator \texttt{PYTHIA 8.2}~\cite{Sjostrand:2014zea} is used to simulate the $d N_{e^+}/d p$ of positron produced by DM annihilation.
The DM particle mass $m_\chi$ and velocity-weighted annihilation cross section $\langle \sigma_\text{ann} v_\text{rel}\rangle$ are  free parameters
to be determined by the AMS-02 data.
In order to take into account of  some uncertainties in the secondary positron flux, such as that from the hadronic interactions, the nuclear enhancement factor from the heavy elements, and the interstellar medium density distribution, we allow the calculated fluxes of secondary positrons to vary global with a free factor $C_{e^+}$, and take it as an extra free parameter.
We perform a Bayesian analysis to the data.  To efficiently explore the parameter space, we adopt the \texttt{MultiNest} sampling algorithm~\cite{ Feroz:2007kg, Feroz:2008xx, Feroz:2013hea}.

The fit results including posterior means, standard deviations, and best-fit values are summarized in Tab.~\ref{tab:best-fit}, and the regions favored by the AMS-02 data in the parameter space $(m_\chi, \langle\sigma_{\text{ann}} v_{\text{rel}}\rangle)$ are shown in the left panel of Fig.~\ref{fig:ams-pos}.
It shows that  $m_\chi \approx 1-1.6\ \rm{TeV}$ with $\langle\sigma_{\text{ann}} v_{\text{rel}}\rangle \approx (3-6)\times 10^{-24}\ \rm{cm^3\ s^{-1}}$ is favored in this configuration.
These results confirm our previous analysis in the heavy mediator limit~\cite{Jin:2014ica}.
We considered several different mediator masses from 1~GeV to 1~TeV. The fit results are found to be insensitive to $m_\phi$, if   $m_\phi\ll m_\chi$. 
In the left panel of Fig.~\ref{fig:ams-pos}, we also show the upper limits from Fermi-LAT observations on dSphs gamma rays, which are derived by performing a likelihood analysis following the Fermi-LAT collaboration~\cite{ Ackermann:2015zua} with heavy mediator mass $\sim 1\ \rm{TeV}$, and constraints from measurement of CMB from Planck~\cite{Aghanim:2018eyx} at $95\%$ C.L..
The details of constraints will be discussed in Sec.~\ref{sec:Fermi} and \ref{sec:CMB}.
It shows that the regions favored by AMS-02 positron data are in strong tension with the constraints from Planck observation on CMB and only marginally compatible with the constraints from Fermi-LAT observations on dSphs gamma rays, which is a known problem for constant DM annihilation cross section.
In the right panel of Fig.~\ref{fig:ams-pos}, we show the calculated CR positron flux with best-fit DM parameters, as well as the secondary positron backgrounds for two different mediator masses. As it can be seen from the figure and also Tab.~\ref{tab:best-fit}, the cases of light mediators predict broader spectra, and are in better agreement with the data. 
%The backgrounds have a very weak dependence on the DM properties due to the variation of best-fit $C_{e^+}$.
%

To have an estimation of  the uncertainties arising from the  CR propagation models, we also perform fits using the ``MIN'' and ``MAX'' propagation models  listed in Tab.~\ref{tab:propagation1}.  We find that changing the propagation models from ``MED'' to ``MIN'' (``MAX'') model can lead to a rescaling of $m_\chi$ and $\langle\sigma_{\text{ann}} v_{\text{rel}}\rangle$ favored by AMS-02 positron data by a factor of $\sim 0.78$ (1.62) and $\sim 0.89$ (1.48), respectively (see appendix~\ref{app} for details). Such a difference does not change the main conclusion of our analysis.

%we compare the fit results of three representative propagation models, ``MIN'', ``MED'', and ``MAX'' models  listed in Tab.~\ref{tab:propagation1}. The results are shown in appendix~\ref{app}.
%As can be seen from the appendix that ,changing the propagation models from ``MED'' to ``MIN'' (``MAX'') model can lead to a rescaling of $m_\chi$ and $\langle\sigma_{\text{ann}} v_{\text{rel}}\rangle$ favored by AMS-02 positron data by a factor of $\sim 0.78$ (1.62) and $\sim 0.89$ (1.48), respectively.

\begin{table}[t]
%\begin{tabular}{c|cc|cc|cc|c}
%\hline \hline
%\multirow{2}{*}{$m_\phi\ (\rm{GeV})$} & \multicolumn{2}{c|}{$\rm{log_{10}}(m_\chi/\rm{GeV})$} & \multicolumn{2}{c|}{$\rm{log_{10}}(\langle\sigma_{\text{ann}} v_{\text{rel}}\rangle/\rm{cm^3\ s^{-1}})$} & \multicolumn{2}{c|}{$C_{e^+}$} & \multirow{2}{*}{$\chi^2/d.o.f$}\\
%\cline{2-7} & (Mean, $\sigma$) & Best-fit & (Mean, $\sigma$) & Best-fit & (Mean, $\sigma$) & Best-fit & \\
%\hline
%1 & $3.08 \pm 0.03$ & 3.08 & $-23.39 \pm 0.04$ & -23.39 & $1.61 \pm 0.03$ & 1.61 & 31.06/32 \\
%100 & $3.08 \pm 0.03$ & 3.08 & $-23.38 \pm 0.04$ & -23.38 & $1.61 \pm 0.03$ & 1.61 & 30.98/32 \\
%500 & $3.11 \pm 0.03$ & 3.11 & $-23.35 \pm 0.04$ & -23.36 & $1.62 \pm 0.03$ & 1.62 & 30.56/32\\
%1000 & $3.15 \pm 0.02$ & 3.14 & $-23.35 \pm 0.03$ & -23.36 & $1.69 \pm 0.02$ & 1.69 & 52.93/32 \\
%\hline \hline
%\end{tabular}
\begin{tabular}{c|cc|cc|cc|c}
	\hline \hline
	\multirow{3}{*}{$m_\phi/\rm{GeV}$} & \multicolumn{2}{c|}{$\rm{log_{10}}(m_\chi/\rm{GeV})$} & \multicolumn{2}{c|}{$\rm{log_{10}}(\langle\sigma_{\text{ann}} v_{\text{rel}}\rangle/\rm{cm^3\ s^{-1}})$} & \multicolumn{2}{c|}{$C_{e^+}$} & \multirow{3}{*}{$\chi^2/\text{d.o.f}$}\\
	\cline{2-7} & \multicolumn{2}{c|}{Prior range: [1, 4]} & \multicolumn{2}{c|}{Prior range: [-26, -21]} & \multicolumn{2}{c|}{Prior range: [0.1, 10]}\\
	\cline{2-7} & (Mean, $\sigma$) & Best-fit & (Mean, $\sigma$) & Best-fit & (Mean, $\sigma$) & Best-fit & \\
	\hline
	1 & $3.08 \pm 0.03$ & 3.08 & $-23.39 \pm 0.04$ & -23.39 & $1.61 \pm 0.03$ & 1.61 & 31.06/32 \\
	100 & $3.08 \pm 0.03$ & 3.08 & $-23.38 \pm 0.04$ & -23.38 & $1.61 \pm 0.03$ & 1.61 & 30.98/32 \\
	500 & $3.11 \pm 0.03$ & 3.11 & $-23.35 \pm 0.04$ & -23.36 & $1.62 \pm 0.03$ & 1.62 & 30.56/32\\
	1000 & $3.15 \pm 0.02$ & 3.14 & $-23.35 \pm 0.03$ & -23.36 & $1.69 \pm 0.02$ & 1.69 & 52.93/32 \\
	\hline \hline
\end{tabular}
\caption{Prior ranges, posterior means, standard deviations and best-fit values of DM mass, annihilation cross section, and the normalization factor for different mediator masses in the ``MED'' propagation model. The values of $\chi^2/\text{d.o.f}$ are also listed as an estimation of the goodness-of-fit.}
\label{tab:best-fit}
\end{table}

\section{$p$-wave Sommerfeld enhancement of dark matter annihilation}
\label{sec:se&re}

The Sommerfeld enhancement of DM annihilation cross section occurs when the annihilating DM particles self-interact through a long-range attractive potential formed by multiple-exchange of light mediator particles~\cite{https://doi.org/10.1002/andp.19314030302}. In this case, the short-distance annihilation cross section can be enhanced as the wave functions of the annihilating particles at origin are distorted from plane wave.
When a fermionic DM particle $\chi$ couples to a light scalar mediator $\phi$, the induced attractive potential is a Yukawa potential $V_Y(\boldsymbol{r})=-\alpha e^{-m_\phi r}/r$, where $m_\phi$ is the mass of the mediator, and $\alpha=g^2/4\pi $ is the coupling constant. The two-body wave function $\Psi(\boldsymbol{r})$ of the annihilating DM particles satisfies the following non-relativistic  Schr\"{o}dinger equation
\begin{equation}
-\frac{1}{m_{\chi}} \nabla^{2} \Psi(\mathbf{r})+V_Y(\mathbf{r}) \Psi(\mathbf{r})=m_{\chi} v^{2} \Psi(\mathbf{r})\ ,
\end{equation}
where $v=v_{\rm{rel}}/2$ is the velocity of DM particles in the center-of-mass frame and $v_{\rm{rel}}$ is the DM relative velocity.
The wave function can be expanded over angular momentum $\ell$, namely, $\Psi(r, \theta)=\sum_{\ell} P_{\ell}(\cos \theta) \chi_{\ell}(r) / r$, where $P_{\ell}$ is the Legendre polynomial and $\chi_{\ell}(r)$ is the radial wave function. Adopting  the dimensionless parameters $\epsilon_v = v/\alpha,\ \epsilon_\phi = m_\phi/(\alpha m_\chi)$ and rescaling the radial coordinate $r$ with $x=\alpha m_\chi r$, the Schr\"{o}dinger equation for the   radial wave functions $\chi_{\ell}(x)$ can be written as
\begin{equation} \label{eq:Schrodinger}
\chi_{\ell}^{\prime \prime} (x)+\left[\epsilon_v^2 - V_\ell(x)- V_Y(x)\right] \chi_{\ell}(x)=0\ ,
\end{equation}
where $V_\ell(x)=\ell(\ell+1)/x^{2}$ is the centrifugal potential, and $V_Y(x)=-e^{-\epsilon_\phi x}/x$ is the Yukawa potential in terms  of $x$.
Eq.~(\ref{eq:Schrodinger}) should be solved  with the following  boundary conditions~\cite{Iengo:2009ni, Cassel:2009wt}
\begin{equation}
\lim _{x \rightarrow 0} \chi_{\ell}(x)=(\epsilon_v x)^{\ell+1} \text {\ , and } \lim _{x \rightarrow \infty} \chi_{\ell}(x) \rightarrow C_\ell \sin \left(\epsilon_v x-\frac{\ell \pi}{2}+\delta_{\ell}\right)\ ,
\end{equation}
where $C_\ell$ is a normalization constant, and $\delta_\ell$ is the phase shift of the $\ell$-th partial wave.
The Sommerfeld enhancement factor for the $\ell$-th partial wave is defined as~\cite{ArkaniHamed:2008qn}
\begin{equation}
S_{\ell}^Y \equiv \lim _{x \rightarrow 0}\left|\frac{\chi_{\ell}(x)} {\chi_{\ell}^{(0)}(x)}\right|^{2}=\left[\frac{(2 \ell+1) ! !}{C_\ell}\right]^{2}\ ,
\end{equation}
where $\chi_{\ell}^{(0)}(x)$ is the $\ell$-th radial wave function which is the solution of the Schr\"{o}dinger equation in the absence of a  potential.

Approximate analytic solutions can be obtained by approximating the Yukawa potential with the Hulth$\acute{e}$n potential~\cite{Cassel:2009wt,Tulin:2013teo}
\begin{equation}
V_Y(x)\approx V_H(x)=-\frac{\delta e^{-\delta x}}{1-e^{-\delta x}}\ ,
\end{equation}
where $\delta = \pi^2 \epsilon_\phi/6$.
In this approximation, the $s$-wave Sommerfeld enhancement factor  $S_{0}^H$  has  the well-known form~\cite{Tulin:2013teo}
\begin{equation}\label{eq:s0}
S_{0}^H=\frac{\pi}{\varepsilon_{v}} \frac{\sinh \left(\frac{2 \pi \varepsilon_{v}}{\pi^{2} \varepsilon_{\phi} / 6}\right)}
{\cosh \left(\frac{2 \pi \varepsilon_{v}}{\pi^{2} \varepsilon_{\phi} / 6}\right)-\cos \left(2 \pi \sqrt{\frac{1}{\pi^{2} \varepsilon_{\phi} / 6}
-\frac{\varepsilon_{v}^{2}}{\left(\pi^{2} \varepsilon_{\phi} / 6\right)^{2}}}\right)}\ .
\end{equation}
The $s$-wave Sommerfeld enhancement factor has two important characters.
One is that the Sommerfeld enhancement saturates at low velocity, once the deBroglie wavelength of the incident particle gets larger than the range of interaction, equivalently, $\epsilon_v \ll \epsilon_\phi$.
The other is the additional resonant enhancement from the threshold bound states developed by the attractive potential for specific values of $\epsilon_\phi \simeq  6/(n^2 \pi^2),\ (n=1,2,3\dots)$.
In the case of $s$-wave annihilation, the centrifugal potential vanishes and the Hulth$\acute{e}$n potential can approximate the Yukawa potential very well.  Thus Eq.~(\ref{eq:s0}) is an excellent approximation of $S^Y_0$, which typically reproduces the numerical results with uncertainties less than $\sim10\%$, and accurately reproduces the resonant behavior~\cite{Feng:2010zp}.

For higher partial waves with $\ell \neq 0$, in order to obtain approximate analytic expressions,  additional approximation has to be applied to the centrifugal potential~\cite{Cassel:2009wt}
\begin{equation}
V_\ell(x)\approx \tilde{V}_{\ell}(x) = \ell(\ell+1) \frac{\delta^{2} e^{-\delta x}}{\left(1-e^{-\delta x}\right)^{2}}\ .
\end{equation}
Under this approximation, the analytic expression of the $p$-wave Sommerfeld enhancement factor is given by~\cite{Tulin:2013teo}
\begin{equation} \label{eq:s1}
S_{1}^H=\frac{(1-\varepsilon_{\phi} \pi^2/6)^2+4\varepsilon_v^2}{(\varepsilon_{\phi} \pi^2/6)^2+4\varepsilon_v^2}S_0^H\ .
\end{equation}
Although this approximation is reasonable at  short distances with $\delta \cdot x \ll 1$, it does not well reproduce the long-distance behavior of $V_\ell(x)$.
\begin{figure}[t]
\includegraphics[width=\textwidth]{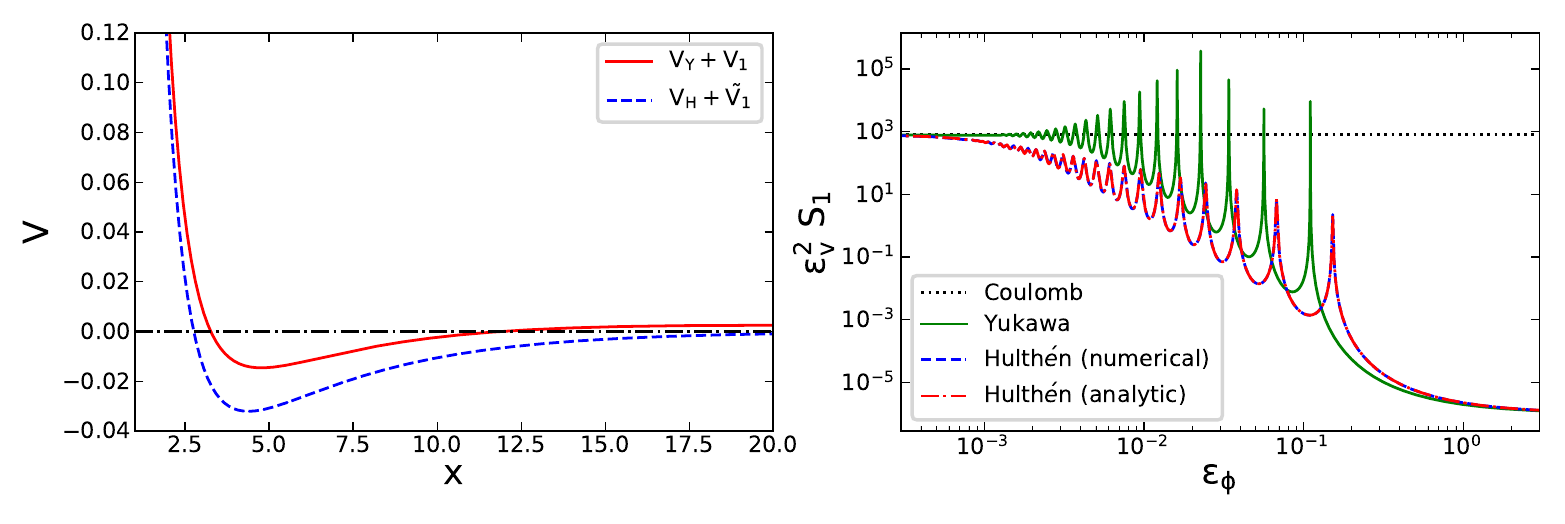}
\caption{
	Left) Comparison between the Yukawa potential plus the centrifugal potential  $V_Y(x)+V_{\ell=1}(x)$ and the Hulth$\acute{e}$n potential plus the approximate centrifugal potential $V_H(x)+\tilde{V}_{\ell=1}(x)$ as a function of $x$ with  $\epsilon_\phi$ fixed at $=6/(2^2 \pi^2)$. 
	Right) Values of the rescaled $p$-wave Sommefeld enhancement factor $\epsilon_v^2 S_1^Y$  as a function of $\epsilon_{\phi}$ at fixed $\epsilon_v=10^{-3}$. The green solid line represents the numerical solution for Yukawa potential, the red dashed line is the approximate analytic result for Hulth$\acute{e}$n potential, the blue dashed line stands for the numerical result for Hulth$\acute{e}$n potential as a comparison. The black dot line corresponds to  the Coulomb limit.}
\label{fig:potential&Sp}
\end{figure}
%
%== Fig.2 left
In the left panel of Fig.~\ref{fig:potential&Sp},  we show the difference between $V_H(x)+\tilde{V}_{\ell=1}(x)$  and $V_Y(x)+V_{\ell=1}(x)$. It can be seen that the difference becomes significant near the force-range of the Yukawa and Hulth$\acute{e}$n potentials, i.e.,  $\sim 1/\epsilon_\phi$. For this reason,  Eq.~(\ref{eq:s1}) is not a reasonable approximation for $p$-wave process due to the mismatch of the two potentials, especially near the resonant regions.
%== Fig.2 right
We  directly calculate the $p$-wave Sommerfeld enhancement factors for Yukawa potential from numerically solving Eq.~(\ref{eq:Schrodinger}).  Since the $p$-wave  annihilation cross section before the enhancement  is proportional to $v_{\rm{rel}}^2$, we show the rescaled enhancement factor  $\epsilon_v^2 S_1^Y$ calculated numerically from the Schr\"{o}dinger equation  and $\epsilon_v^2 S_1^H$ calculated using Eq.~(\ref{eq:s1}) in the right panel of Fig.~\ref{fig:potential&Sp}. As the figure shows, the difference between the two method is significant. In general, the direct numerical calculation for the Yukawa potential gives significantly larger enhancement factor and different locations of the resonance, which may have important phenomenological consequences.
For an estimation of the accuracy of the numerical solutions, we numerically solve the Sommerfeld enhancement factors for the Hulth$\acute{e}$n potential case and compare them with the analytic solutions. We find an excellent agreement between the two methods, which is also shown in Fig.~\ref{fig:potential&Sp}.
%
%
%Both of them approach to the Coulomb limit $\pi(1+4\epsilon_v^2)/(4\epsilon_v(1-e^{-\pi/\epsilon_v}))$~\cite{ArkaniHamed:2008qn} at $\epsilon_\phi \ll \epsilon_v$, which is the product of $\epsilon_v^2$ and $p$-wave Sommerfeld enhancement factor for Coulomb potential ($V_C(x)=-1/x$).
%
Therefore, we  shall adopt the numerical values of $p$-wave Sommerfeld enhancement factor $S_1^Y$ in this work.
%
%$S^Y_1$ have the same morphologies of resonance and saturation with $S^H_1$,
%although there exist significant discrepancy between them.
%

%TODO 1.1
	The fact that the Hulth$\acute{e}$n-potential approximation  cannot well reproduce  the $p$-wave Sommerfeld enhancement factors induced by the Yukawa potential may have important phenomenological implications. 
	For the  Hulth$\acute{e}$n potential,  as shown in Eq.~(\ref{eq:s1}), the values of $\epsilon_\phi$ at the resonance points are the same for both the $s$- and $p$-wave cases. 
	Our direct numerical calculations indicate that  this is in general {\it not true} for the the Yukawa potential. 
	In the generic case, DM annihilation can involve both $s$- and $p$-wave processes.
	Due to the different locations of the  resonance points, 
	the $s$- and $p$-wave Sommerfeld enhancement will not reach the resonance point simultaneously. 
 	Thus it is possible that for a given parameter set, only the $p$-wave or $s$-wave  Sommerfeld enhancement is significant.

\begin{figure}[t]
\includegraphics[width=\textwidth]{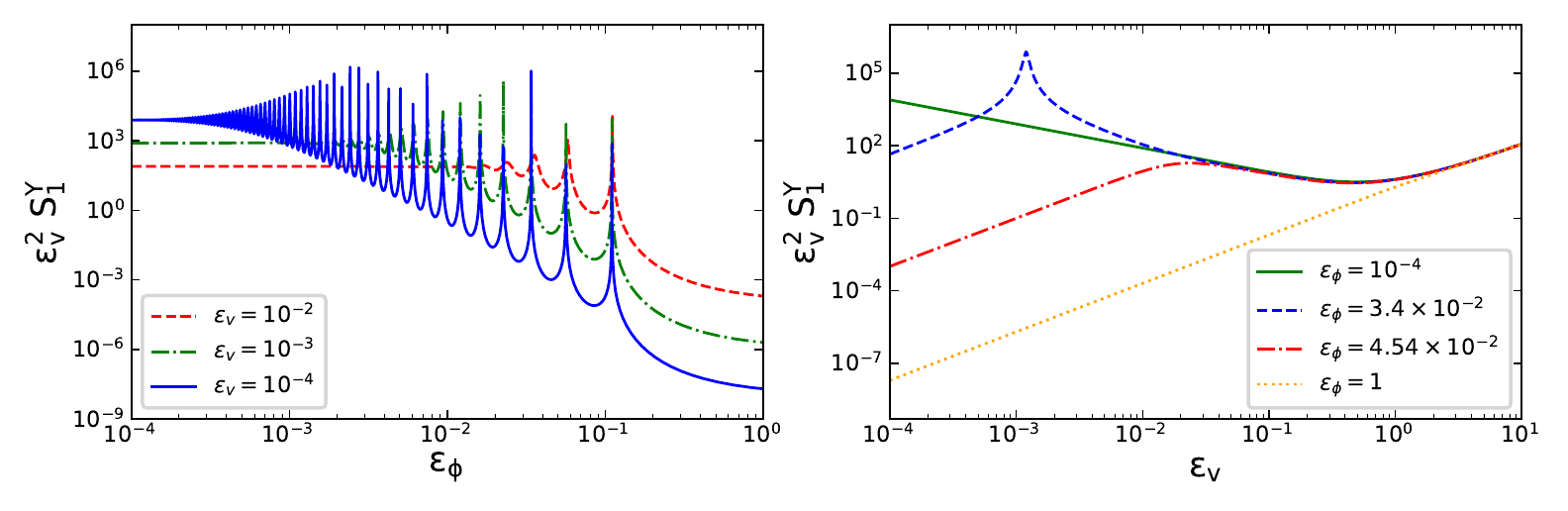}
\caption{Left)  $p$-wave Sommefeld enhacement factor $\epsilon_v^2S_1^Y$ as a function of $\epsilon_\phi$ with $\epsilon_v=10^{-2},\ 10^{-3}$, and $10^{-4}$, respectively.
Right) $\epsilon_v^2S_1^Y$ as a function of $\epsilon_v$ with $\epsilon_\phi=10^{-4},\ 3.4\times 10^{-2},\ 4.54\times 10^{-2}$, and $1.0$, respectively.}
\label{fig:Sp_ep_ev}
\end{figure}

Another major difference from the $s$-wave case is that the $p$-wave Sommerfeld-enhanced annihilation cross section can be both velocity-enhanced (for $\epsilon_v\gg \epsilon_\phi$) and velocity-suppressed (for $\epsilon_v\ll \epsilon_\phi$), which is also  important in  phenomenology.
In the left panel of Fig.~\ref{fig:Sp_ep_ev}, we show the  value of $\epsilon_v^2 S_1^Y$  as a function of the dimensionless parameter $\epsilon_\phi$ for three typical value of $\epsilon_v=10^{-2},10^{-3}$, and $10^{-4}$, respectively. It shows that $\epsilon_v^2 S_1^Y$ is higher (lower) at lower $\epsilon_v$ for $\epsilon_\phi \ll \epsilon_v$ ($\epsilon_\phi \gg \epsilon_v$). 
In the right panel of Fig.~\ref{fig:Sp_ep_ev}, we show the dependence of $\epsilon_v^2 S_1^Y$ on dimensionless parameters $\epsilon_v$ for four typical values of $\epsilon_\phi=10^{-4},3.4\times 10^{-2},4.54\times 10^{-2}$, and $1.0$. 
The value of $3.4\times 10^{-2}$ ($4.54\times 10^{-2}$) is a representative value for 
$\epsilon_\phi$ which is on (off) the resonance region.
%($\epsilon_\phi=3.4\times 10^{-2},\ 4.54\times 10^{-2}$ are parameter values on and away from resonances).
%
For $\epsilon_\phi\gtrsim 1$ or $\epsilon_v\gtrsim 1$, there is no Sommerfeld enhancement, $S_1^Y\sim 1$, thus $\epsilon_v^2 S_1^Y$ scales as $\epsilon_v^2$.
For $\epsilon_\phi \lesssim \epsilon_v \lesssim 1$, $S_1^Y$ approaches the Coulomb limit, hence $\epsilon_v^2S_1^Y$ scales as $1/\epsilon_v$.
For $\epsilon_v \lesssim \epsilon_\phi \lesssim 1$, when the effect of the Sommerfeld enhancement is not saturated, $\epsilon_v^2 S_1^Y$ scales as $1/\epsilon_v^2$ near the resonance regions and $1/\epsilon_v$ away from the resonance regions.
Once $S_1^Y$ has saturated ($\epsilon_\phi \gg \epsilon_v$), $S_1^Y$ remains the same and $\epsilon_v^2 S_1^Y$ scales as $\epsilon_v^2$.
It is possible that the complicated velocity dependence may allow the $p$-wave Sommerfeld enhancement  to accommodate the data of DM relic density, AMS-02 positron flux, gamma rays of dSphs, and CMB simultaneously.

\section{Constraints from thermal relic density}\label{sec:relic-density}

The time (thermal) evolution of the DM number density $n$ is governed by the Boltzmann equation~\cite{Bertone:2004pz}
\begin{equation} \label{eq:Bol}
\frac{dn}{dt}+3H(t)n=-\left<\sigma_{\rm{ann}} v_{\rm{rel}}\right>\left(n^2-n_{eq}^2\right)\ ,
\end{equation}
where $H(t)$ is the Hubble parameter, $n_{eq}$ is the DM number density in equilibrium.
Changing variables from $t\to x=m_\chi/T$, $n\to Y=n/s$ and $n_{eq}\to Y_{eq}=n_{eq}/s$, where $T$ is the temperature of thermal bath and $s$ is the entropy density, the Boltzmann equation  can be rewritten as~\cite{Kolb:1988aj}
\begin{equation}
\label{eq:Bolzmann}
\frac{d Y}{d x}= - \sqrt{\frac{\pi}{45}}m_{\rm{pl}} m_{\chi}g_{\ast}^{-1/2} g_{\ast s} 
x^{-2}\left<\sigma_{\rm{ann}} v_{\rm{rel}}\right>\left(Y^2-Y_{eq}^2\right)\ ,
\end{equation}
where $m_{\rm{pl}}\simeq 1.22\times 10^{19}\ \mathrm{GeV}$ is the Planck mass scale, 
$g_{\ast}$ and $g_{\ast s}$ are the effective relativistic degrees of freedom 
for energy and entropy density, respectively~\cite{Steigman:2012nb}.
In a $p$-wave Sommerfeld enhanced DM annihilation model, the annihilation cross section is $\sigma_{\rm{ann}} v_{\rm{rel}}=(\sigma_{\rm{ann}} v_{\rm{rel}})_0\times S_1$, where $(\sigma_{\rm{ann}} v_{\rm{rel}})_0 = bv_{\rm{rel}}^2$ is the short-distance cross section and $b$ is a global factor.
If the DM particles are in thermal equilibrium, the  velocity distribution  function $f(v)$ of DM particles is the Maxwell-Boltzmann distribution 
\begin{equation}\label{eq:mb}
	f(v)=\left(\frac{1}{\pi v_0^2}\right)^{3/2} e^{-\frac{v^2}{v_0^2}}\ ,
\end{equation}
with $v_0$ the most probable velocity.
The velocity-weighted annihilation cross section with the Sommerfeld enhancement is given by
\begin{equation} 
	\langle \sigma_{\rm{ann}} v_{\rm{rel}}\rangle =\frac{b}{v_0^3} \sqrt{\frac{2}{\pi}} \int_0^{+\infty} \mathrm{d}v_{\rm{rel}} e^{-\frac{v_{\rm{rel}}^2}{2v_0^2}} v_{\rm{rel}}^4 S_1(v_{\rm{rel}}/2\alpha, \epsilon_\phi)\ .
	\label{eq:cs}
\end{equation}
The velocity-weighted DM annihilation cross section is assumed to be in the form of Eq.~(\ref{eq:cs}) with $v_0=\sqrt{2T_\chi/m_\chi}$, where $T_\chi$ is the  temperature of DM particles. If the DM particles are in  equilibrium with the thermal bath, $T_\chi=T$.

\begin{figure}[t]
\includegraphics[width=0.7\textwidth]{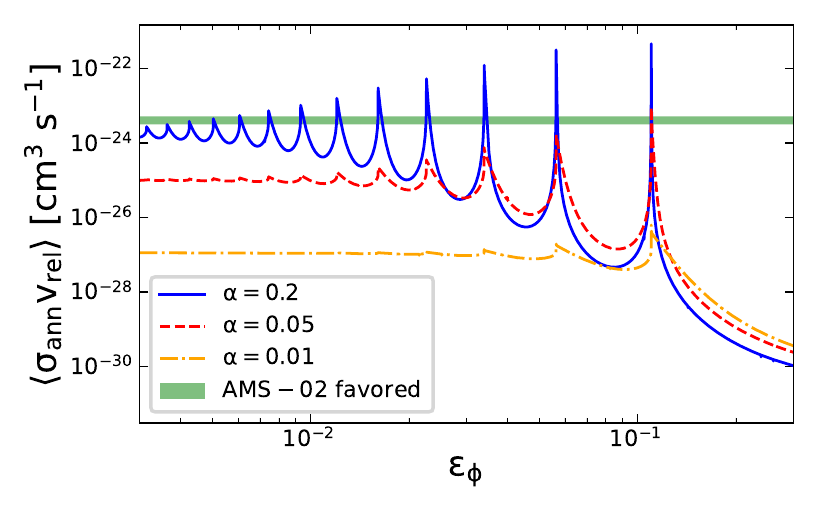}
\caption{$\langle \sigma_{\rm{ann}} v_{\rm{rel}}\rangle$ calculated from Eq.~(\ref{eq:cs}) with $v_0=220\ \rm{km\ s^{-1}}$ as a function of $\epsilon_\phi$ for $\alpha=0.01,\ 0.05$, and $0.2$, where DM mass is fixed at the best-fit value of AMS-02 positron data $m_\chi=1.2\ \rm{TeV}$ and $b$ is derived from the correct DM relic density. The green area is favored by AMS-02 positron data~\cite{Aguilar:2019owu} at $95\%$ CL with $m_\phi=1\ \rm{GeV}$ for $4\mu$ channel in ``MED'' propagation model.}
\label{fig:BF}
\end{figure}

The relic abundance of DM particles can be written as~\cite{Chen:2013bi}
\begin{equation}
	\Omega_{\chi} h^2 \approx 2.76 \times 10^8 Y(x_\infty)\left(\frac{m_\chi}{\rm{GeV}}\right)\ ,
\end{equation}
where $x_\infty$ is the value of $x$ in the present day.
The observed value of DM relic abundance from Planck is $\Omega_{\chi}h^2\simeq0.1193$~\cite{Aghanim:2018eyx}. 
We numerically solve the Boltzmann equation of Eq.~(\ref{eq:Bolzmann}) with the Sommerfeld-enhanced DM annihilation cross section. During the calculation we have assumed that the mediator particles are in thermal equilibrium at freeze-out of DM particles. 
As shown in~\cite{Chen:2013bi}, although the effect of kinetic decoupling in $s$-wave process reduces DM relic abundance significantly, the kinetic decoupling effect in Sommerfeld enhanced $p$-wave annihilation is negligible, if the DM freeze-out temperature $T_f$ is higher than kinetic decoupling temperature $T_f/T_{\rm{kd}}\gtrsim \mathcal{O}(1)$, which is easily satisfied.   We thus do not consider the kinetic decoupling in this work, namely, the relation  $T_\chi\approx T$ is assumed.

%
%To determine the DM relic density, we must numerically solve the Boltzmann equation.
%%
%Sommerfeld enhancement can give rise to the kinetic decoupling effect~\cite{Hryczuk:2010zi, Zavala:2009mi, Dent:2009bv, Chen:2013bi}.
%%
%Kinetic decoupling of DM occurs at $T_{\rm{kd}}$, when the momentum transfer rate through elastic scattering with mediator particles drops below the Hubble expansion rate.
%%
%We assume that the mediator particles are in thermal equilibrium at freeze-out of DM.
%%
%DM temperature $T_\chi$ equals the temperature of thermal bath $T$ before kinetic decoupling, and equals $T^2/T_{\rm{kd}}$ after kinetic decoupling~\cite{Chen:2001jz, Hofmann:2001bi, Bringmann:2006mu}.
%%
%As shown in~\cite{Chen:2013bi}, kinetic decoupling effect in Sommerfeld enhanced $p$-wave annihilation is negligible if DM freeze-out temperature $T_f$ is higher than kinetic decoupling temperature $T_f/T_{\rm{kd}}\gtrsim \mathcal{O}(1)$, although this effect in $s$-wave process reduces DM relic abundance significantly.
%%
%We will not consider the kinetic decoupling here, namely, $T_\chi\approx T$.

With the requirement that the observed DM relic abundance $\Omega_{\chi} h^2$ must be reproduced,
the value of  $b$ in the expression of DM annihilation cross section can be determined for  given parameters  $\alpha$, $m_\chi$ and $m_\phi$.
Using the determined value of $b$, the $p$-wave annihilation cross section $\langle \sigma_{\rm{ann}} v_{\rm{rel}}\rangle$ can be predicted from Eq.~(\ref{eq:cs}), for a given velocity distribution characterized by the most probable velocity $v_0$.
In the Galactic halo $v_0=v_{\rm{halo}}\approx 220\ \rm{km\ s^{-1}}$.
%
%The velocity distribution of the DM particles, which can make a significant contributions to local high-energy CR positron spectra, can be describe by Maxwell-Boltzmann distribution Eq.~(\ref{eq:mb}) with the most probable velocity $v_0=v_{\rm{halo}}\approx 220\ \rm{km\ s^{-1}}$.
%
%
%
%In order to compare with the DM explanation for the AMS-02 positron observations, we convert the constraint on $b$ to $\langle \sigma_{\rm{ann}} v_{\rm{rel}}\rangle_{v_0=v_{\rm{halo}}}$ through Eq.~(\ref{eq:cs}).
%
In Fig.~\ref{fig:BF}, we show the predicted value of  $\langle \sigma_{\rm{ann}} v_{\rm{rel}}\rangle_{v_{\rm{halo}}}$ as a function of $\epsilon_\phi$ after the constraints from the DM thermal relic density, for three typical values of $\alpha = 0.2,\ 0.05$, and $0.01$, respectively.
The DM mass is fixed at the best-fit value of AMS-02 positron data, $m_\chi = 1.2\ \rm{TeV}$ and 
the mediator mass is fixed at $m_\phi=1$~GeV.
The figure shows that in general a large coupling constant $\alpha \gtrsim 0.1$ is needed to achieve a large enough DM annihilation cross section to explain the AMS-02 positron data.
A large $\alpha$ may lead to large annihilation cross section for the process of $\chi\chi\to 2\phi$ which will also be subjected to the constraints from DM thermal relic density~\cite{Feng:2009hw,Feng:2010zp}.
This kind of constraint is, however,  model dependent.  It is possible to have $\alpha$ at the scale of $\mathcal{O}(10^{-1})$ in some well-motivated DM models. 
For instance, in the case when the interaction Lagrangian has the form $\mathcal{L}_{\rm{int}}=- g{\chi}\chi\phi/2-\mu \phi ^3/3!$, 
where $g=\sqrt{4\pi \alpha}$, the  annihilation ${\chi}\chi \to \phi\phi$ is a $p$-wave process with tree-level cross section~\cite{Liu:2013vha}
\begin{equation}
  (\sigma_{\rm{ann}} v_{\rm{rel}})_0 = \kappa\frac{3\pi \alpha^2}{8m_\chi^2}v^2_{\rm{rel}}\ 
\end{equation}
with $\kappa = 1-5\xi/18+\xi^2/48$, where $\xi=\mu /(m_\chi g)$~\cite{Chen:2013bi}.
The global factor $b$ has the form $b=3\pi k \alpha^2/(8m_\chi^2)$.
Using an approximate analytic solution to the Boltzmann equation Eq.~(\ref{eq:Bolzmann}) in which 
the Sommerfeld enhancement is neglected due to the high velocity of DM particles at freeze-out,
the requirement from the correct thermal cross section is~\cite{Bertone:2004pz, Jungman:1995df}
\begin{equation}
\frac{6 b}{x_f} \approx 3\times 10^{-26}\ \rm{cm^3\ s^{-1}}\ ,
\end{equation}
where $x_f\sim 20$ is the value of $x$ at freeze-out.
Hence  the scale of the coupling constant $\alpha$ is estimated as 
\begin{equation}
\alpha \sim 0.25\left(\frac{0.1}{k}\right)^{\frac{1}{2}} \left(\frac{m_\chi}{1\ \rm{TeV}}\right)\ .
\end{equation}
In the case of cubic self-interaction, the minimum value of $k$ is $2/27$, which can lead to $\alpha \sim 0.3 \left(m_\chi/1\ \rm{TeV}\right)$.
We thus set set the benchmark value of $\alpha\approx 0.2$ in this work.

\section{Constraints from Fermi-LAT gamma-ray observations of Dwarf spheroidal satellite galaxies}\label{sec:Fermi}

The dSphs are particularly promising targets for DM indirect detection due to their 
low diffuse Galactic gamma-ray backgrounds and high DM density~\cite{Grcevich:2009gt}.
The gamma-ray flux from DM self-annihilation within an energy range ($E_{\min}$, $E_{\max}$) from a solid angle $\Delta\Omega$ is given by
\begin{equation} 
\label{eq:flux}
\begin{aligned}
\Phi_\gamma(\Delta \Omega, E_{\min}, E_{\max})=&\frac{1}{8\pi m_{\chi}^{2}} \int_{E_{\min}}^{E_{\max}} \frac{dN_\gamma}{dE_\gamma} dE_\gamma \\
&\times \int_{\Delta \Omega} d \Omega \int_{\text{l.o.s}} dl \int d^{3} \boldsymbol{v}_1 \int d^{3} \boldsymbol{v}_2 f\left(\boldsymbol{r}, \boldsymbol{v}_1\right) f\left(\boldsymbol{r}, \boldsymbol{v}_2\right) \left(\sigma_{\rm{ann}} v_{\rm{rel}}\right)\ ,
\end{aligned}
\end{equation}
where $f\left(\boldsymbol{r}, \boldsymbol{v}\right)$ is the DM phase-space distribution function.
%
%In the simplest case where $\sigma_{\rm{ann}} v_{\rm{rel}}$ is a constant, Eq.~(\ref{eq:flux}) can be written as~\cite{Ackermann:2015zua}
In the  cases where the velocity-dependence of $\sigma_{\rm{ann}} v_{\rm{rel}}$ can be factorized out, the gamma-ray flux can be written as
\begin{equation}\label{eq:gamma_flux} 
	\Phi_\gamma(\Delta \Omega, E_{\min}, E_{\max}) =
 	\frac{C}{8\pi m_{\chi}^{2}} 
 	\int_{E_{\min}}^{E_{\max}} 
 	\frac{dN_\gamma}{dE_\gamma}{dE_\gamma}\times J ,
 %\int_{\Delta \Omega} d \Omega \int_{{l.o.s}} dl \rho^{2}}\ ,
\end{equation}
%
%
%\begin{equation} 
%\label{eq:J0}
%\Phi_\gamma(\Delta \Omega, E_{\min}, E_{\max}) =\underbrace{\frac{\sigma_{\rm{ann}} v_{\rm{rel}}}{8\pi m_{\chi}^{2}} \int_{E_{\min}}^{E_{\max}} \frac{dN_\gamma}{dE_\gamma} dE_\gamma}_{\rm{pp}} \times \underbrace{ \int_{\Delta \Omega} d \Omega \int_{{l.o.s}} dl \rho^{2}}_{\mathrm{J_0}}\ ,
%\end{equation}
%%
%where the $\rm{pp}$ term only depends on the particle physics properties.
%
where $C$ is a velocity-independent part of the annihilation cross section and the $J$-factor contains   the information of DM density distribution and the velocity-dependent part of the cross section.
% == J0
For instance, in the simplest case where  $\sigma_{\rm{ann}} v_{\rm{rel}}$  is velocity independent. $C=\sigma_{\rm{ann}} v_{\rm{rel}}$ and the corresponding $J$-factor is 
\begin{equation} \label{eq:J0}
J_0=\int_{\Delta \Omega} d \Omega \int_{\text{l.o.s}}  \rho^{2}(\mathbf{r}) dl\ ,
\end{equation}
The $J_0$-factor encapsulates purely the astrophysical information, which contains a line-of-sight (l.o.s) integral through the square of DM distribution $\rho(\mathbf{r})$ over a slolid angle $\Delta \Omega$.
In the case of $s$-wave Sommerfeld-enhanced DM annihilation,  $C=(\sigma_{\rm{ann}} v_{\rm{rel}})_0$
and the $J$-factor becomes
\begin{equation}
	J_s=\int_{\Delta \Omega} d \Omega \int_{\text{l.o.s}} dl \int d^{3} \boldsymbol{v}_1 \int d^{3} \boldsymbol{v}_2 f\left(\boldsymbol{r}, \boldsymbol{v}_1\right) f\left(\boldsymbol{r}, \boldsymbol{v}_2\right) S_0(v_{\text{rel}}) .
	\label{eq:js}
\end{equation}
The integration over the DM velocity should be cut off at the escape velocity $v_{\text{esc}}$ for a given dSph.
Similarly,  in the case of $p$-wave Sommerfeld-enhanced DM annihilation,  $C=b$ and the corresponding $J$-factor is 
\begin{equation} \label{eq:Jp}
 		J_p=\int_{\Delta \Omega} d \Omega \int_{l.o.s} dl \int d^{3} \boldsymbol{v}_1 \int d^{3} \boldsymbol{v}_2 f\left(\boldsymbol{r}, \boldsymbol{v}_1\right) f\left(\boldsymbol{r}, \boldsymbol{v}_2\right) v_{\text{rel}}^2 S_1(v_{\text{rel}} ) .
 \end{equation}

In this section, we  investigate how the $p$-wave Sommerfeld enhancement affects the $J$-factors for dSphs and derive the constraints on DM annihilation from Fermi-LAT  data on gamma rays from dSphs.
%
%Sommerfeld effect can enhance the $J$-factor and change their relative order, thus can affect the upper limits of DM annihilation cross section from Fermi-LAT dSphs gamma-ray observations.
%
We will consider 15 dSphs in the combined analysis of Fermi-LAT collaboration~\cite{Ackermann:2015zua} which are listed in Table.~\ref{tab:js}.
They are selected from 18 dSphs with kinematically determined $J$-factors.
Three of these 18 dSphs, Canes Venatici I, Leo I and Ursa Major I are excluded to ensure statistical independence between observations, since they overlap with other dSphs.

\subsection{DM velocity distribution within dSphs}

In the case of velocity-dependent DM annihilation cross sections, the gamma-ray flux arising from DM annihilation depends also on the DM velocity distribution within dSphs.
To determine the DM velocity distribution function, we adopt a simple assumption that the gravitation potential of dSphs is spherically-symmetric  and the velocity distribution of DM particles is isotropic. 
In this case, the isotropic distribution function of DM particles with a given energy density profile $\rho_{\chi}(r)$ in dSphs can be determined by the Eddington's formula~\cite{2008gady.book.....B}
\begin{equation} \label{eq:frv}
f_{\chi}(\epsilon)=\frac{1}{\sqrt{8} \pi^{2}} \int_{\epsilon}^{0} 
\frac{d^{2} \rho}{d \Psi^{2}} \frac{d \Psi}{\sqrt{\Psi-\epsilon}}\ ,
\end{equation}
where $\Psi$ is the spherical-symmetric gravitational potential, and $\epsilon$ is the gravitational binding energy per mass of a DM particle. 
The function $f_\chi(\epsilon)$ is essentially  the DM phase-space distribution $f_\chi(r, v)$, as $\epsilon=v^2/2+\Psi(r)$ is implicitly a function of both $v$ and $r$.
In this work, the density profile of DM particles in dSphs is assumed 
to be the Navarro-Frenk-White (NFW) profile~\cite{Navarro:1996gj}
\begin{equation}
\rho_{\chi}(r)=\frac{\rho_{s}}{\left(r / r_{s}\right)\left(1+r / r_{s}\right)^{2}}\ ,
\end{equation} 
where the parameters $\rho_{s}$ and $r_{s}$ are the reference density and radius, respectively. 
The two parameters can be determined by the maximum circular velocity $V_{\max }$ and the radius of maximum circular velocity $R_{V_{\max }}$ of the dSph through the relations
\begin{equation}\label{eq:nfwp}
r_{s} =\frac{R_{V_{\max }}}{2.163}\ \text{, and}\ 
\rho_{s} =\frac{4.625}{4 \pi G} \left(\frac{V_{\max}}{r_{s}}\right)^{2}.
\end{equation}
Thus, the DM distribution function $f_{\chi}(r,v)$ depends only on the astrophysical parameters $\left(V_{\max }, R_{V_{\max }}\right)$,
which can be determined by the observation of the average stellar line-of-sight velocity dispersion of each dSph.

\begin{figure}[t]
\includegraphics[width=0.7\textwidth]{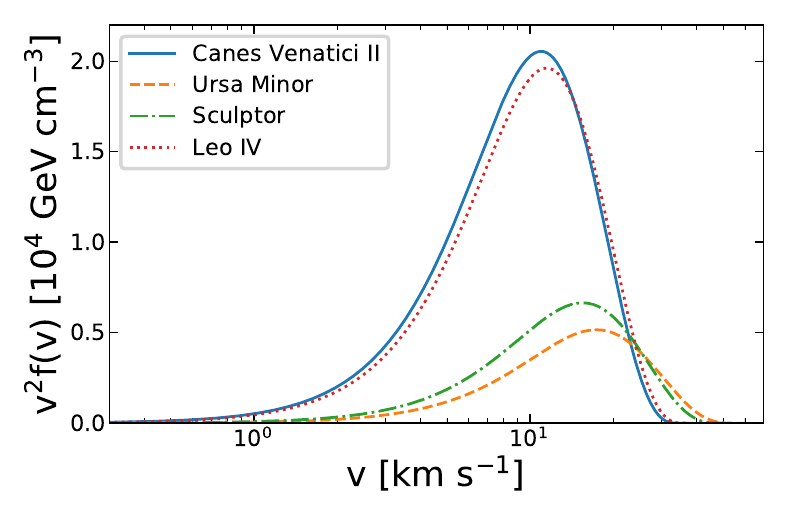}
\caption{Velocity distribution functions $v^2f(v)$ evaluated at radius  $r=r_s/2$  using the Eddington's formula~\cite{2008gady.book.....B} for  a selection of four dSphs Canes Venatici II, Ursa Minor, Sculptor, and Leo IV.}
\label{fig:dis_vel}
\end{figure}

In Fig.~\ref{fig:dis_vel}, we show the velocity distribution rescaled by $v^2$ calculated from Eq.~({\ref{eq:frv}}) $v^2f(r, v)$ at radius $r=r_s/2$ as a function of $v$ for a selection of four typical dSphs, Canes Venatici II, Ursa Minor, Sculptor, and Leo IV selected from the 15 dSphs. 
The values of $\left(V_{\max }, R_{V_{\max }}\right)$ for these dSphs are taken from Ref.~\cite{Martinez:2013els}, and the NFW profile parameter $\left(\rho_{s}, r_{s}\right)$ are determined by Eq.~(\ref{eq:nfwp}).
%
%The results are based on the simplified assumption of NFW DM density profile and isotropic velocity distribution.
%
%Contrary to the predictions of cusp density profile from N-body numerical simulations,
 Note that the measured rotation curves of some dSphs suggests the  cored DM profile~\cite{Kuhlen:2012ft,Diemand:2009bm,deBlok:2009sp}.
Extending the calculations for  non-NFW profiles is straightforward. It only requires modifying the relation of Eq.~(\ref{eq:nfwp}) accordingly.
%
%is similar within the framework of isotropic distribution, except that the relation between the profile parameters with $V_{\max}$ and $R_{V_{\max}}$ should be modified.
%
The value of $V_{\max}$ and $R_{V_{\max}}$ of each dSph with other profiles, such as cored NFW, Burkert, and Einasto profile can be found  in Ref.~\cite{Martinez:2013els}. 
%
%It may be necessary to go beyond Eddington's formula when considering the anisotropic velocity dispersion.

\subsection{$p$-wave Sommerfeld-enhanced $J$-factor}

Using the isotropic phase-space distribution of DM particles in dSphs of Eq.~(\ref{eq:frv}), the expression of $J_p$ can be written as
\begin{equation}
\begin{aligned} J_p &=\frac{32 \pi^{3}}{D^{2}} \int_{0}^{r_{\max}} r^{2} d r 
\int_{0}^{v_{\rm{esc}}(r)} v_{1}^{2} f_\chi\left(r, v_{1}\right) d v_{1} \int_{0}^{v_{\rm{esc}}(r)} v_{2}^{2} f_\chi\left(r, v_{2}\right) d v_{2} \\ 
& \times \int_{0}^{\pi} \sin \theta S_1\left(\sqrt{v_{1}^{2}+v_{2}^{2}-2 v_{1} v_{2} \cos \theta }/ 2\alpha, \epsilon_\phi\right) (v_{1}^{2}+v_{2}^{2}-2 v_{1} v_{2} \cos \theta ) d \theta\ ,\end{aligned}
\label{eq:jp}
\end{equation}
where $D$ is the distance from the Solar system to the center of the dSphs under consideration, whose values are taken from Ref.~\cite{Ackermann:2015zua}. $v_{\mathrm{esc}}(r)=\sqrt{-2 \Psi(r)}$ is the escape velocity at radius $r$.
The integral over solid angle is performed over a circular region with a solid angle of $\Delta\Omega\sim 2.4\times 10^{-4}\ \rm{sr}$, namely, angular radius of $\beta = 0.5^{\circ}$ which corresponds to the maximal radius $r_{\rm{max}}=D \cdot \rm{sin}\beta$. In the assumption of NFW profile of DM distribution in dSphs, the set of DM halo scale radii $r_s$ span a range of subtended angles between $0.1^\circ$ and $0.4^\circ$~\cite{Ackermann:2013yva}. Since $\Delta\Omega$ is large enough to essentially encompass the entire region of the dSphs in which there is significant dark matter annihilation, the $J$-factor can be expressed in terms of an integral over the radial distance from the center of the dSphs, instead of an integral over the line of sight.
The value of $J_p$ depends on five parameters: two of them are related to particle physics: $\alpha$ and $\epsilon_\phi$.  Another three are astrophysical parameters: $D$, $V_{\max}$ and $R_{V_{\max}}$.

\begin{figure}[t]
\includegraphics[width=0.7\textwidth]{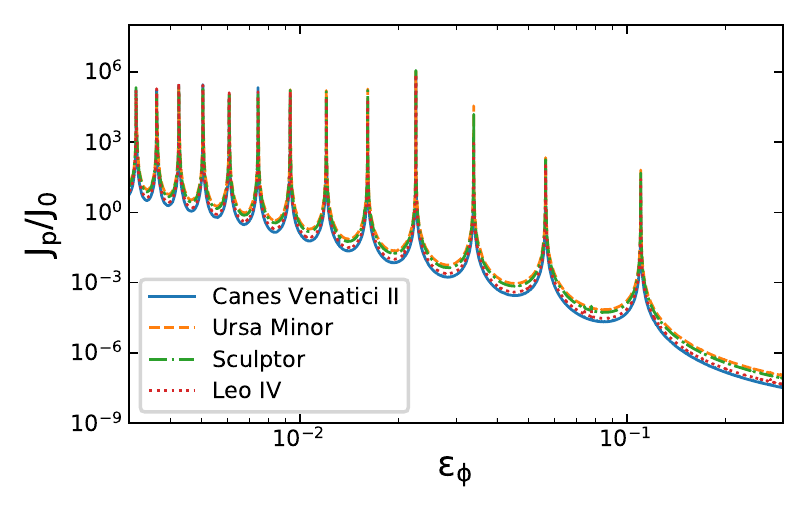}
\caption{The ratio of $J_p/J_0$ as a function of $\epsilon_\phi$ for the four dSphs Canes Venatici II, Ursa Minor, Sculptor, and Leo IV.  The value of the coupling strength is fixed at $\alpha=0.2$.
}  
\label{fig:Js}
\end{figure}

\begin{table}
\begin{tabular}{ c c c c c c }
\hline \hline
Name & $J_p(0.005)[\times 10^{20}]$ & $J_p(0.006)[\times 10^{19}]$ & $J_p(0.009)[\times 10^{18}]$ & $J_p(0.05)[\times 10^{15}]$ & $J_0[\times 10^{18}]$ \\
\hline
Bootes I & $3.1$ & $13.3$ & $5.7$ & $4.7$ & $6.3$ \\
Canes Venatici II & $0.3$ & $1.1$ & $0.4$ & $0.3$ & $0.8$ \\
Carina & $0.5$ & $2.2$ & $0.9$ & $0.7$ & $1.3$ \\
Coma Berenices & $3.4$ & $13.9$ & $5.5$ & $4.6$ & $10.0$\\
Draco & $5.0$ & $22.2$ & $9.7$ & $8.2$ & $6.3$ \\
Fornax & $1.1$ & $5.0$ & $2.3$ & $1.9$ & $1.6$ \\
Hercules & $0.5$ & $1.9$ & $0.8$ & $0.7$ & $1.3$ \\
Leo II & $0.2$ & $0.7$ & $0.3$ & $0.2$ & $0.4$ \\
Leo IV & $0.3$ & $1.4$ & $0.6$ & $0.5$ & $0.8$ \\
Sculptor & $2.8$ & $11.9$ & $5.2$ & $4.4$ & $4.0$ \\
Segue 1 & $11.4$ & $46.1$ & $17.7$ & $14.9$ & $31.6$ \\
Sextans & $1.4$ & $5.7$ & $2.4$ & $2.0$ & $2.5$ \\
Ursa Major II & $8.8$ & $35.8$ & $15.1$ & $12.5$ & $20.0$ \\
Ursa Minor & $5.1$ & $22.8$ & $10.3$ & $9.0$ & $6.3$ \\
Willman 1 & $4.4$ & $18.0$ & $7.3$ & $5.9$ & $12.6$ \\
\hline \hline
\end{tabular}
\caption{$J_p(\epsilon_\phi)$ (in unit of  $\mathrm{GeV^2\ cm^{-5}}$) of 15 dSphs at $\epsilon_\phi=0.005,\ 0.006,\ 0.009$, and $0.05$ with $\alpha=0.2$. The last column shows the values of $J_0$ from~\cite{ Ackermann:2015zua}.
}
\label{tab:js}
\end{table}

In Fig.~\ref{fig:Js}, we show the ratio between the $p$-wave Sommerfeld-enhanced $J$-factor $J_p$ and the  normal $J$-factor $J_0$ calculated in~\cite{ Ackermann:2015zua}  as a function of $\epsilon_\phi$ for the four dSphs considered in Fig.~\ref{fig:dis_vel}. In Table.~\ref{tab:js}, we list the values of $J_p$ at four typical values of $\epsilon_\phi=0.005,\ 0.006,\ 0.009$, and $0.05$ for all the fifteen dSphs.
%and the last column is $J_0$ taken from Ref.~\cite{Ackermann:2015zua}. 
%
Similar to the $s$-wave Sommerfeld-enhanced $J$-factor $J_s$ calculated in Ref.~\cite{Boddy:2017vpe,Lu:2017jrh}, $J_p$ has the resonant behavior at certain values of $\epsilon_\phi$ due to the Sommerfeld enhancement factor in Eq.~(\ref{eq:jp}).
Near the resonant regions, $J_p$ can be a few  orders of magnitude larger than $J_0$.
Unlike $J_s$ which is always greater than $J_0$, the value of $J_p$ can be suppressed at some $\epsilon_\phi$, where the Sommerfeld enhancement is subdominant compared with the velocity suppression factor $v_{\text{rel}}^2$. 
In the limit of a heavy mediator, there is no Sommerfeld enhancement.
Thus $J_s/J_0$ approaches to $1$, while $J_p/J_0$ approaches to $v_{\rm{dSphs}}^2$, where $v_{\rm{dSphs}}\approx 10~\mathrm{km\ s}^{-1}$ is the DM typical velocity in dSphs.
In the case of  $p$-wave Sommerfeld enhancement, the dSphs which have  similar $J_0$ factors may have quite different values of $J_p$. For instance  Bootes I and Draco have nearly the same $J_0$ factor of $6.3\times 10^{18}~\mathrm{GeV^2 cm^{-5}}$, but the corresponding $J_p$ factors at $\epsilon_\phi=0.005$ are $3.1\times 10^{20}~\mathrm{GeV^2 cm^{-5}}$ and 
$5.0\times 10^{20}~\mathrm{GeV^2 cm^{-5}}$, respectively. This changes the  relative importance of $J_p$ among the dSphs in their contribution to the gamma-ray flux. For instance, Ursa Minor and Draco become more important in $J_p$ than they are in $J_0$. However,  Segue 1 (Leo II) remains to be  the brightest (faintest) dSph among all the fifteen dSphs.

$J_p$ calculated in this work is based on the assumption of isotropic velocity distribution and NFW profile of DM in dSphs. Extending the calculation of $J_p$ to other DM distributions is straightforward.
Due to the low dispersion of DM velocity in dSphs, $J_p$ can be approximately  as $J_p \sim v_{\rm{dSphs}}^2 S_1(v_{\rm{dSphs}}) J_0$. It has been shown  that $J_0$ is fairly insensitive to the assumed DM density profile~\cite{Martinez:2009jh, Strigari:2013iaa}, hence $J_p$ is expected  to weakly dependent of the DM profile in dSphs.

\subsection{Constraints on DM annihilation cross section from Fermi-LAT dSphs gamma-ray data}
\label{sec:Fer_analysis}

We place constraints on the DM annihilation cross sections using  six years of Fermi-LAT  PASS8 data on gamma rays from the fifteen dSphs listed in Tab.~\ref{tab:js}. The gamma-ray fluxes are measured in the energy range of $500\ \mathrm{MeV}-500\ \mathrm{GeV}$, and are binned into 24 logarithmically-spaced bins. %(marked with $j,\ j=1,2,...24$).
The likelihood function for an individual dSph with index  $i$ is given by~\cite{ Ackermann:2015zua}
\begin{equation}
\mathcal{L}_{i}\left(\boldsymbol{\mu}, \hat{\boldsymbol{\theta}_{i}} \mid \mathcal{D}_{i}\right)=\prod_{j=1}^{24} \mathcal{L}_{ij}\left(\boldsymbol{\mu}, \hat{\boldsymbol{\theta}_{i}} \mid \mathcal{D}_{ij}\right)\ ,
\end{equation}
where $\mathcal{L}_{ij}\left(\boldsymbol{\mu}, \hat{\boldsymbol{\theta}_{i}} \mid \mathcal{D}_{ij}\right)$ is the likelihood function for the $i$-th dSph in the $j$-th energy bin, 
$\boldsymbol{\mu}$ stands for the parameters of the DM model,
$\hat{\boldsymbol{\theta}_{i}}$ are the best-fit values of the nuisance parameters such as the flux normalizations of background gamma-ray sources, and $\mathcal{D}_{ij}$ represent the observed gamma-ray fluxes.
The likelihood as a function of gamma-ray flux is explicitly given by the Fermi-LAT collaboration in~\cite{Fermi-LAT}.
To determine the constraint on the parameters of a given DM model $\boldsymbol{\mu}$, we adopt the standard delta-log-likelihood approach. We calculate the quantity $\Delta\chi^2(\boldsymbol{\mu})$ as follows
\begin{equation}
\Delta\chi^2(\boldsymbol{\mu})=-2 \ln \left(\frac{\mathcal{L}_i\left(\boldsymbol{\mu}, \hat{\boldsymbol{\theta}_i} \mid \mathcal{D}_i\right)}{\mathcal{L}_i(\hat{\boldsymbol{\mu}}, \hat{\boldsymbol{\theta}_i} \mid \mathcal{D}_i)}\right)\ ,
\end{equation}
where $\hat{\boldsymbol{\mu}}$ are the values which maximize $\mathcal{L}_i\left(\boldsymbol{\mu}, \hat{\boldsymbol{\theta}_i} \mid \mathcal{D}_i\right)$.
The upper limits on the DM parameter $\boldsymbol{\mu}$ at a given $m_\chi$ from Fermi-LAT gamma-ray data on dSph $i$ at $95\%$ C.L. are determined by requiring $\Delta\chi^2 \leq 2.71$, if $\boldsymbol{\mu}$ is a one-dimensional quantity.
For a  combined analysis for all the dSphs, one can replace the individual likelihood function of dSph $i$ $\mathcal{L}_i\left(\boldsymbol{\mu}, \hat{\boldsymbol{\theta}_i} \mid \mathcal{D}_i\right)$ with a joint likelihood function $\mathcal{L}\left(\boldsymbol{\mu}, \hat{\boldsymbol{\theta}} \mid \mathcal{D}\right)=\prod_{i=1}^{15} \mathcal{L}_{i}\left(\boldsymbol{\mu}, \hat{\boldsymbol{\theta}_{i}}\mid \mathcal{D}_{i}\right)$ in the above equation.
Note that statistical uncertainties in $V_{\rm{max}}$ and $R_{V_{\rm{max}}}$  may affect the $J$-factor determination~\cite{Martinez:2013els}. These uncertainties can give rise to the uncertainties of DM density profile in dSphs through Eq.~(\ref{eq:nfwp}), and then the $J$-factor (with and without Sommerfeld enhancement) through Eq.~(\ref{eq:J0}-\ref{eq:Jp}). For DM annihilation models with constant cross section, the Fermi-LAT collaboration has incorporated this part of uncertainties in the $J_0$ determination as a nuisance parameter with an extra likelihood function in Gaussian form~\cite{Ackermann:2015zua}. As shown in~\cite{Lu:2017jrh}, the upper limits of DM annihilation cross section can be modified by at most $2\%-5\%$ when this part of astrophysical uncertainties  is not taken into account. For simplicity, we thus do not consider the uncertainties in $V_{\rm{max}}$ and $R_{V_{\rm{max}}}$  in this work.

%== constant xsection J0
%
%For DM annihilation models with constant cross section, the discrepancy of upper limits on DM annihilation %cross section with and without considering the statistical uncertainty in the $J$-factor $J_0$ determination is %at most $2\% - 5\%$~\cite{Lu:2017jrh}, due to the small value of $J$-factor likelihood. We thus calculate the %constraints on DM annihilation without taking into account the $J$-factor likelihood.
%
%
For DM annihilation models with constant (velocity-independent) cross sections, the parameter set of DM model $\boldsymbol{\mu}$ is simply $\langle\sigma_{\rm{ann}} v_{\rm{rel}}\rangle$. The gamma-ray flux produced by DM annihilation in dSphs is proportional to $J_0$ and can be calculated using Eq.~(\ref{eq:J0}).
In this case, the upper limits on the DM annihilation cross section at $95\%$ C.L. are determined through the combined analysis using the joint likelihood $\mathcal{L}$. The results are shown in the left panel of  Fig.~\ref{fig:ams-pos}. It can be seen that the resulting constraints are only marginally compatible with the region favored by the  AMS-02 CR positron data.
%
%The Fermi-LAT collaboration has shown that  the differences in the resulting upper limits with and without considering the statistical uncertainty in  $J_0$ determination is at most $2\% - 5\%$~\cite{Lu:2017jrh}. We thus calculate the constraints  without taking into account the $J$-factor likelihood.

For the $s$-wave Sommerfeld-enhanced DM annihilation models, the upper limits on DM annihilation cross section from Fermi-LAT gamma-ray data on dSphs have been determined in Ref.~\cite{Lu:2017jrh}. The calculation is similar to the constant cross section case, except that the J-factor $J_0$ should be replaced by $J_s$.
Since the $s$-wave Sommerfeld enhancement factor $S_0$ increases monotonically with decreasing DM velocity, for $s$-wave Sommerfeld-enhanced DM annihilation, the  corresponding  upper limits  are lower  roughly by a factor of $S_0(v_{\rm{dSphs}})/S_0(v_{\rm{halo}})>1$. Thus it becomes even more difficult to reconcile  the AMS-02 CR positron data and Fermi-LAT gamma-ray data in this scenario.

\begin{figure}[t]
\includegraphics[width=\textwidth]{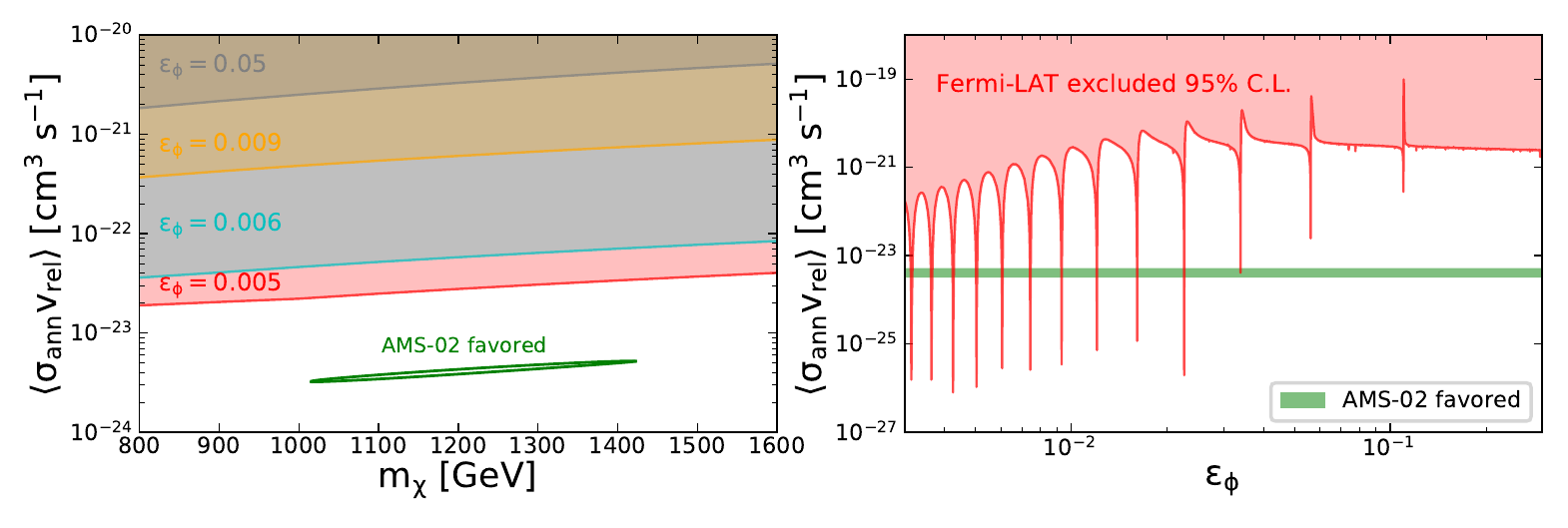}
\caption{
	Left) Upper limits on DM annihilation cross section at $95\%$ C.L. as a function of DM particle mass $m_\chi$  derived from a combined analysis on the  Fermi-LAT  gamma-ray data of the 15 dSphs listed in Table.~\ref{tab:js}~\cite{Ackermann:2015zua} for four different values of $\epsilon_\phi$ . The DM annihilation channel is $\chi\chi\to 2\phi\to 4\mu$.
	The value of the coupling strength is fixed at $\alpha=0.2$.
	Right) The same upper limits as a function of $\epsilon_\phi$, where $m_\chi$ is fixed at  $1.2\ \mathrm{TeV}$. 
	In both panels, the green regions are favored by AMS-02 CR positron data~\cite{Aguilar:2019owu} at $95\%$ C.L..}
\label{fig:fermi}
\end{figure}

In $p$-wave Sommerfeld-enhanced DM annihilation models, the situation can be quite different. 
In this case, the parameter $\boldsymbol{\mu}$ corresponds to  the global factor $b$.
With the Sommerfeld-enhanced $J$-factor $J_p$ calculated in the previous subsection, it is straight forward to  calculate the gamma-ray flux produced by the DM annihilation in dSphs from Eq.~(\ref{eq:gamma_flux}).
We derive the upper limits on $b$ at $95\%$ C.L. from the Fermi-LAT gamma-ray data as a function of $m_\chi$  for some  specific values of $\epsilon_\phi$ using the delta-log-likelihood approach.
%
%== Fig.7 (left)
To facilitate the comparison with the  AMS-02 data, we convert the obtained upper limits on the value of  $b$ to that on $\langle \sigma_{\rm{ann}} v_{\rm{rel}}\rangle$ through Eq.~(\ref{eq:cs}) with $v_0=v_{\rm{halo}}$.
In the left panel of Fig.~\ref{fig:fermi}, we show the upper limits for four typical values of $\epsilon_\phi=0.005,\ 0.006,\ 0.009$, and $0.05$ respectively. It can be seen that for these parameters, the limits are compatible with the region  favored by the AMS-02 data.
%== Fig.7 (right)
In the right panel of Fig.~\ref{fig:fermi}, we show the upper limits as a function of $\epsilon_\phi$ with $m_\chi$ fixed at the best-fit value of 1.2~TeV from the AMS-02 data. In the regions very close to the resonance,  the factor $J_p$ is significantly enhanced and the constraints from the Fermi-LAT gamma-ray data become very stringent, which exclude some  of the AMS-02 favored parameter region.
Note that, although the successful explanation of the AMS-02 positron data and the DM thermal relic density also requires that $\epsilon_\phi$ should be close to the resonance regions, such  regions  in general do not overlap with that excluded by the Fermi-LAT gamma-ray data. Thus a consistent explanation for all the observable is still possible.
%
%However, in the regions away from the resonances, the upper limits  from Fermi-LAT  gamma-ray data are typically $\sim 10^{-21}\ \rm{cm^3\ s^{-1}}$, which are about two orders of magnitude higher than the cross section favored by AMS-02 positron data.
%
%In these regions, the tension between the AMS-02 positron excess and Fermi-LAT  gamma-rays constraints  are much weaker than that in models with $s$-wave Sommerfeld-enhanced or constant annihilation cross section.

\section{Constraints from CMB}\label{sec:CMB}

The recombination history of the Universe can be modified by energy injection into gas, photon-baryon plasma and background radiation from DM annihilation, which can lead to modifications in the temperature and polarization power spectra of CMB~\cite{Chen:2003gz,Padmanabhan:2005es}.
Thus, the measurement on the anisotropy of CMB  can be used to constrain the nature of DM particles~\cite{Galli:2009zc,Slatyer:2009yq,Finkbeiner:2011dx}.
Recently, the Planck collaboration reported upper limits on the DM annihilation cross section as 
$f_{\rm{eff}}\langle \sigma_{\rm{ann}} v_{\rm{rel}} \rangle/m_{\chi}\leq 3.2 \times 10^{-28}\ \mathrm{cm^3\ s^{-1}\ GeV^{-1}}$ at $95\%$~C.L.~\cite{Aghanim:2018eyx}, 
where $f_{\rm{eff}}$ is an effective factor describing the fraction of energy  from DM annihilation ionizing and heating the intergalactic medium.
Using the value of $f_{\rm{eff}}\approx 0.16$ for $4\mu$  annihilation channel~\cite{Madhavacheril:2013cna} and assuming that $\langle \sigma_{\rm{ann}} v_{\rm{rel}} \rangle$  is velocity independent, we show in Fig.~\ref{fig:ams-pos} the $95\%$~C.L. upper limits on DM annihilation cross section as a function of DM particle mass from the Planck data. 
The constraints are apparently in strong tension with the AMS-02 favored parameter region if the cross section is velocity independent.
For the $s$-wave Sommerfeld-enhanced DM annihilation, the constraints are expected to be even more stringent.
In the case of $p$-wave Sommerfeld-enhanced DM annihilation, due to the extremely low DM averaged-velocity of $\sim 1\ \mathrm{m\cdot s^{-1}}$ in the epoch of recombination, the annihilation cross section is expected to be strongly suppressed by the $v^2$ factor, which leads to rather weak constraints from the CMB measurements.

\section{Combined constraints from AMS-02, relic dencity, Ferimi-LAT and Planck}
\label{sec:combine}

\begin{figure}[t]
\includegraphics[width=0.49\textwidth]{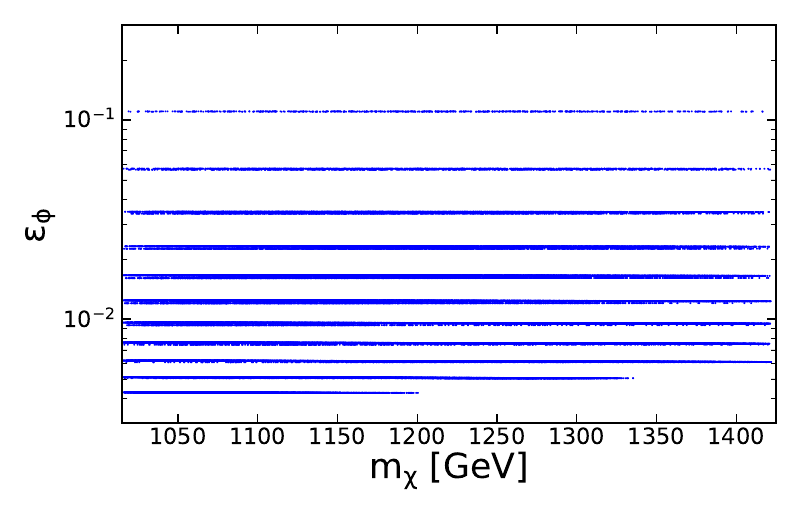}
\includegraphics[width=0.49\textwidth]{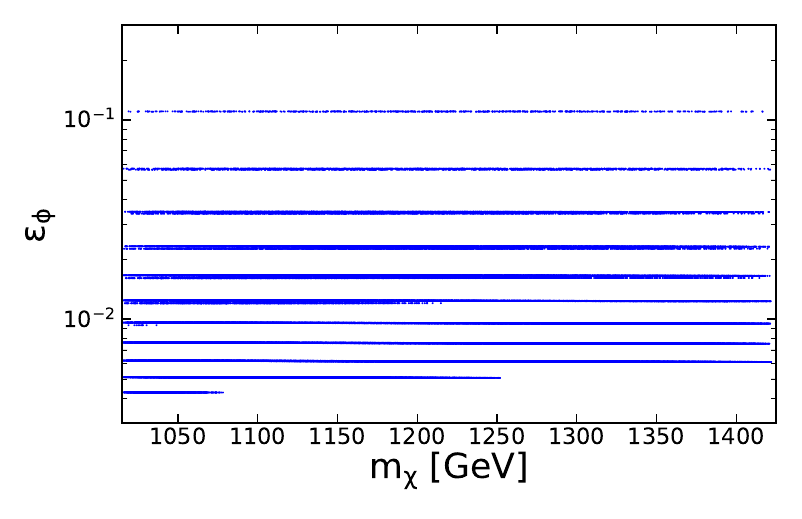}
\caption{
	Left) Parameter regions  which can simultaneously account for the AMS-02 CR positron excess~\cite{Aguilar:2019owu} and  DM relic density~\cite{Aghanim:2018eyx} in the $(m_\chi, \epsilon_\phi)$ plane. The value of the coupling strength is fixed at $\alpha=0.2$.
	Right) Parameter regions  which can simultaneously account for the AMS-02 CR positron excess and  DM relic density, and  still allowed by the constraints from gamma-ray data of 15 nearby dSphs measured by  Fermi-LAT~\cite{Ackermann:2015zua} and  CMB measured by Planck~\cite{Aghanim:2018eyx}.	
}\label{fig:scan}
\end{figure}

\begin{figure}[t]
\includegraphics[width=\textwidth]{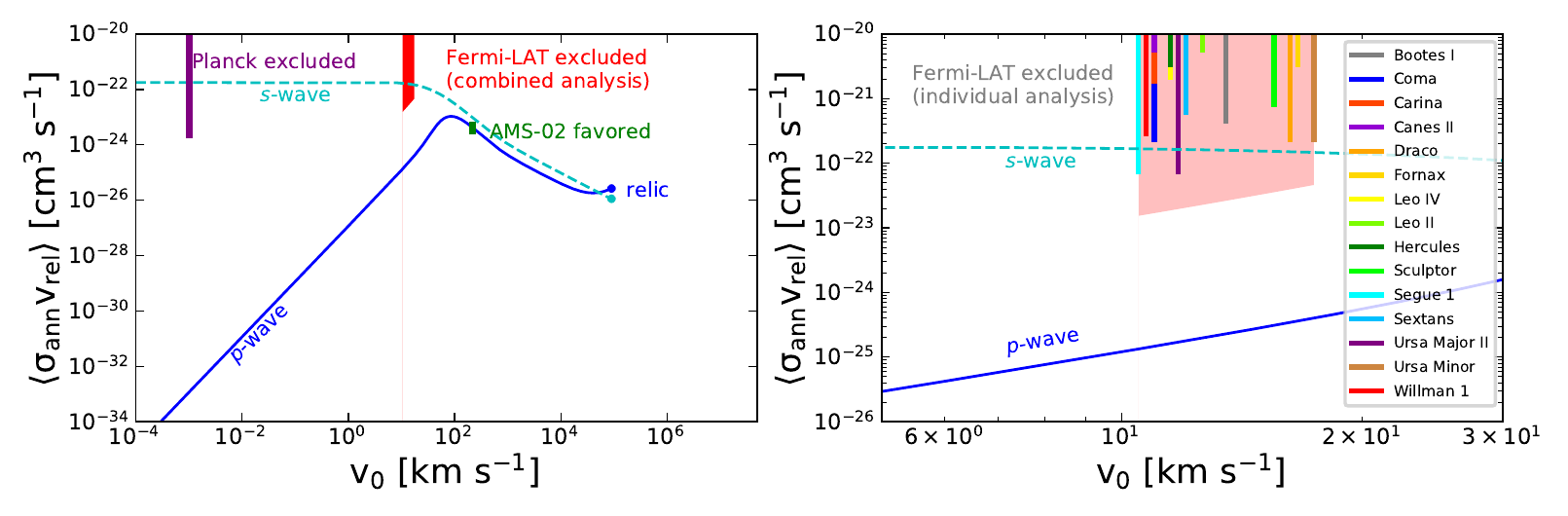}
\caption{
 Left) p-wave Sommerfeld-enhanced DM annihilation cross section as a function of the DM most probable velocity $v_0$ with the global factor $b$ fixed by the DM thermal relic density (blue solid). The DM annihilation channel is $\chi\chi\to 2\phi\to 4\mu$ with $m_\chi=1.2\ \mathrm{TeV}$, $m_\phi=1.8\ \mathrm{GeV}$ and $\alpha=0.2$.
For comparison, the case of an s-wave Sommerfeld-enhanced DM annihilation cross section 
calculated from Eq.~(\ref{eq:s0}) is also shown (cyan dashed).
The green region stands for the cross section favored by AMS-02 positron data~\cite{Aguilar:2019owu} with $v_0\sim 220\ \mathrm{km\ s^{-1}}$.
The vertical region  correspond to  the upper limits at $95\%$~C.L. from a joint analysis of the 15 dSphs measured by Fermi-LAT~\cite{Ackermann:2015zua} with typical  $v_0\sim 10-20~\mathrm{km\ s^{-1}}$.
The purple vertical line indicates the cross section excluded by Planck CMB observations~\cite{Aghanim:2018eyx} at $95\%$ C.L.  with typical $v_0\sim 1\ \mathrm{m\ s^{-1}}$.
Right) Constraints from individual dSphs. The vertical lines with different colors correspond to  the upper limits from Fermi-LAT gamma-ray data of individual dSphs~\cite{Ackermann:2015zua} at $95\%$~C.L., where $v_0$ are taken to be the most probable velocity at $r=r_s/2$ for each dSph.  The shaded region is excluded by the  joint analysis of all the 15 dSphs.
}\label{fig:csv0}
\end{figure}

In order to find the proper parameter space which allows for consistent explanations for the four type of measurements: i) AMS-02 positron excess, ii) DM thermal relic density, iii)  gamma rays of dSphs and iv)  CMB in the $p$-wave Sommerfeld-enhaced DM annihilation, we perform a scan in the two-dimensional parameter space $(m_\chi, \epsilon_\phi)$ with the coupling $\alpha$ fixed at 0.2. The allowed parameters are those satisfying all the constraints at $95\%$~C.L.. 
The allowed parameter regions are shown in Fig.~\ref{fig:scan}. It can be seen that only the discrete regions with $\epsilon_\phi$ being close to the resonant values are allowed by the data of AMS-02 positron excess and DM relic density together. Some of regions are further excluded by the Fermi-LAT gamma-ray data as shown  in the right panel of  Fig.~\ref{fig:scan}. We find that the CMB data from Planck impose almost no additional constraints in the $p$-wave Sommerfeld-enhanced DM annihilation, which is expected from the strong velocity suppression in this processes.
%TODO 3
	Due to the rescaling factor of $\sim 0.89\ (1.48)$ of DM annihilation cross section favored by AMS-02 positron excess in the ``MIN'' (``MAX'') propagation model relative to ``MED'' propagation model, the final parameter regions which can account for both the AMS-02 positron excess and the correct DM relic density while being in agreement with dSphs gamma-ray and CMB constraints would be away from (close to) resonances slightly, and the relevant DM mass is rescaled by a factor of $\sim 0.78\ (1.62)$.

The parameters which can successfully survive all the considered constraints correspond to the  non-trivial velocity dependence of the DM annihilation cross section. For illustration purpose,  we show in Fig.~\ref{fig:csv0} the variation of $\langle \sigma_{\rm{ann}} v_{\rm{rel}} \rangle$ with the velocity $v_0$, for a typical parameters set of  $m_\chi=1.2\ \mathrm{TeV}$, $m_\phi=1.8\ \mathrm{GeV}$ and $\alpha=0.2$.
% relic 
The cross section is fixed at $\sim 3\times 10^{-26}~\mathrm{cm^3s^{-1}}$
by the relic density at $v_0\sim c/3\approx 10^5~\mathrm{km\cdot s^{-1}}$.
% AMS
When the velocity/temperature decreases, the cross section increases due to the Sommerfeld enhancement, and reaches $\sim 3\times 10^{-24}~\mathrm{cm^3s^{-1}}$ for $v_0\sim 220~\mathrm{km\cdot s^{-1}}$ which is large enough to account for the AMS-02 positron excess. 
% dSphs
The cross section continues to increase and reaches a maximal value at $v_0\sim 90~\mathrm{km\cdot s^{-1}}$, then starts to decrease towards lower $v_0$. 
At  $v_0\sim 10~\mathrm{km\cdot s^{-1}}$, the cross section becomes smaller 
than  the constraints of both the combined and individual constraint (right panel of Fig.~\ref{fig:csv0}) from the dSph gamma-ray data measured by Fermi-LAT.
% CMB
Finally at $v_0\sim 1~\mathrm{m\cdot s^{-1}}$, the cross section becomes extremely small such that the CMB cannot impose any constraints.
% compare with  s-wave
In Fig.~\ref{fig:csv0}, we also show the velocity dependence for the  case of $s$-wave Sommerfeld enhancement with the same DM and mediator mass and the cross section at high temperature fixed by the DM thermal relic density. It can be seen that although the $s$-wave Sommerfeld enhancement can explain the CR positron excess, it in general predicts too large cross sections at low velocities, which is in strong tensions with the dSphs gamma-ray data and the measurements of CMB.

\section{Constraints from bounds-state formation}\label{BSF}
% BSF overview
So far we have consider the scenario where the DM particles annihilate  purely through $p$-wave.
In some DM models, there exist additional  $s$-wave annihilation channels 
which are subdominant at freeze out, 
but  gradually becomes important when the temperature is lower
due to the Sommerfeld enhancement effect.
A known example is the bound state formation (BSF) process. 
Let us consider a dark sector consisting a fermionic DM particle $\chi$ and a light scalar mediator $\phi$. 
Due to the same long range force, 
DM particles at low velocity can form bound states $(\chi\bar\chi)$ through 
emitting the mediator $\phi$%
~\cite{Petraki:2015hla,An:2016gad,Cirelli:2016rnw,Petraki:2016cnz}.
Although the DM  annihilation into two mediators is dominated by $p$-wave process
(assuming the interaction conserves parity),
the BSF process $\chi \bar\chi \to (\chi\bar\chi)+\phi$  occurs dominantly through $s$-wave.
The cross section for the $s$-wave BSF process  is very small at high temperature,
but can be dominant at extremely low temperature, 
as  the $p$-wave  DM annihilation cross section is eventually velocity suppressed. 
Due to the $s$-wave  BSF process, this type of model will be stringently  constrained by  the data of CMB~\cite{An:2016kie}.

The cross section for the  dominant monopole transition into the $s$-wave bound state in the limit $v_{\text{rel}}\ll m_\phi/m_\chi$ is given by~\cite{An:2016kie}
\begin{equation}\label{eq:bfs_cs}
	(\sigma_{\rm{bsf}} v_{\rm{rel}}) ^{\text{mono}}_{n, \ell=0}=\frac{2^6\pi^3\alpha^4e^{-4n}(L^1_{n-1}(4n))^2}{9n^3m_\chi^2 \epsilon_\phi \rm{sin}^2(\pi/\sqrt{\epsilon_\phi})}\ ,
\end{equation}
where $L$ is the associated Laguerre polynomial, $n$ and $\ell$ are the principal and angular momentum quantum numbers of the bound state, respectively.
In Eq.~(\ref{eq:bfs_cs}) the spin degrees of freedom for the DM particle is neglected.
Including it will lead to a global factor of 1/4 from spin average of each  state with quantum number $n$ and $\ell$.
In the dominant ground state ($n=1$) formation, 
when the DM particle is a Dirac fermion, there exist two ground states, spin-singlet and spin-triplet with quantum numbers $J^{PC}=0^{-+}$ and $1^{--}$, respectively.
%where $J,\ P,\ C$ are the total spin, parity, and charge parity, respectively.
The $1^{--}$ state, once formed, is stable if the interaction is charge-parity conserving.
The $0^{-+}$ state, on the other hand, can decay to mediators.
If the DM particle is a Majorana fermion (i.e. $\chi=\bar\chi$), the ground state two-DM-particle system can only be in a spin-singlet. 
As a result, only 1/4 of total DM annihilation can take place via the monopole transition into the ground state.
Due to the very short lifetime of the spin-singlet ground state 
compared with the cosmological time scale in the recombination epoch, 
the energy injection rate from DM annihilation through the bound state channel is assumed to be
proportional to the cross section of the spin-singlet ground state formation~\cite{An:2016kie}.

% difference: p-wave, s-wave shift
%
% results

\begin{figure}[thb!]
	\includegraphics[width=0.49\textwidth]{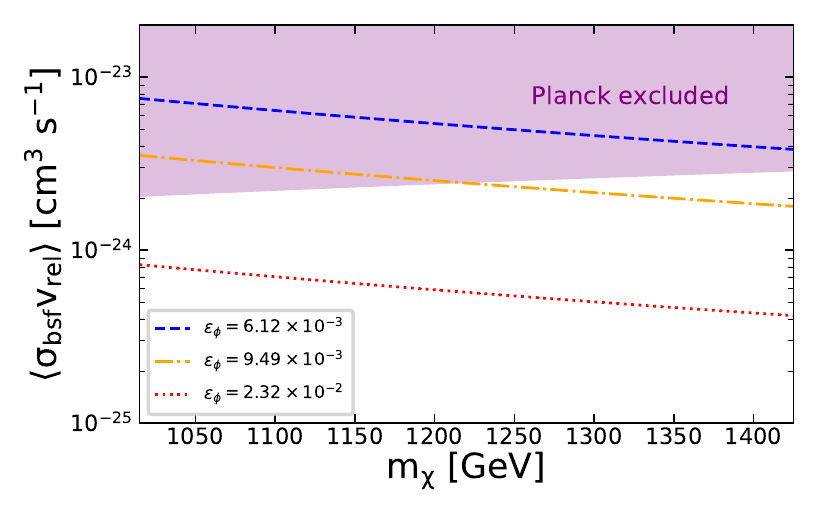}
	\includegraphics[width=0.49\textwidth]{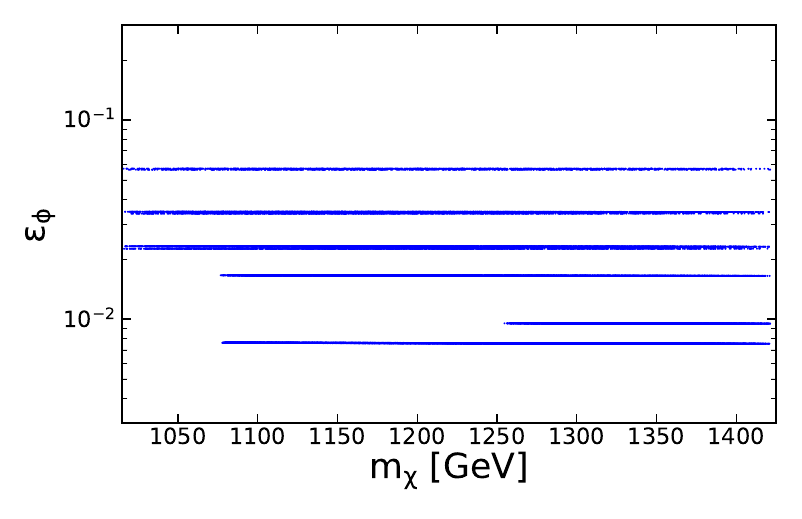}
	\caption{
		Left) The spin-singlet ground state formation cross section as a function of $m_\chi$ at $\epsilon_\phi=6.12\times 10^{-3}, 9.49\times 10^{-3}, 2.32\times 10^{-2}$. The value of $\alpha$ is fixed at 0.2. The shaded region is excluded by CMB constraints from Planck~\cite{Aghanim:2018eyx}.
		Right) Parameter regions in the ($m_\chi, \epsilon_\phi$) plane which can simultaneously account for the AMS-02 CR positron excess~\cite{Aguilar:2019owu} and the correct DM relic density~\cite{Aghanim:2018eyx} while being in agreement with dSphs gamma-ray observations~\cite{Ackermann:2015zua} and CMB constraints from the $s$-wave  bound state formation~\cite{Aghanim:2018eyx}.
	}\label{fig:bsf}
\end{figure}

In the models with  BSF, part of the previously obtained allowed parameter space 
can be further excluded by the CMB data.
In Fig.~\ref{fig:scan}, the allowed parameter points are mainly concentrated in the regions close to resonance of $p$-wave Sommerfeld enhancement.
We take three typical value of $\epsilon_\phi=6.12\times 10^{-3}, 9.49\times 10^{-3}, 2.32\times 10^{-2}$ from the allowed regions in Fig.~\ref{fig:scan}, and show the spin-singlet ground state formation cross section as a function of $m_\chi$ with $\alpha=0.2$ in the left panel of  Fig.~\ref{fig:bsf}.  The upper limits from  the CMB data are also shown for comparison.
It can be seen that  the points with  $\epsilon_\phi=6.12\times 10^{-3}$ are totally excluded by the CMB data 
in the whole DM mass range favored by the AMS-02 data. 
A portion of the DM mass range are excluded for  $\epsilon_\phi=9.49\times 10^{-3}$, 
and the bound state formation cross section at $\epsilon_\phi=2.32\times 10^{-2}$ is still compatible with CMB constraints in the entire DM mass range.
% right panel of Fig.10
%
In the right panel of Fig.~\ref{fig:bsf}, we show the parameter regions surviving the constraints from BSF formation. 
It can be seen that a significant portion of the parameter space can be excluded by introducing this type of processes.
However, in the generic case, the CMB data cannot exclude all the parameter space, because the resonance points for the  $s$- and $p$-wave Sommerfeld enhancement are in general different, as already discussed in the previous section.

\section{Conclusions}\label{sec:conclusion}
In summary, $p$-wave Sommerfeld-enhanced DM annihilation can have complicated velocity dependences compared with that for the $s$-wave case.
We have considered a scenario of  Sommerfeld-enhanced $p$-wave DM annihilation where the DM annihilation cross section can be enhanced at certain velocities but eventually be highly suppressed when the DM velocity approaches zero.
We have performed a systematic calculation of  the velocity-dependent J-factors for the Sommerfeld-enhanced  $p$-wave DM annihilation  for  15 nearby dSphs, and derived constraints from the gamma-ray data of these dSphs from Fermi-LAT.
We have shown that there are parameter regions where it can simultaneously   account for   the CR positron excess,  DM relic density,  gamma-ray data of dSphs from Fermi-LAT, and  CMB measurements from Planck.
Our results show that the CR positron excess can still be consistently explained within the WIMP scenario framework.

\acknowledgements
We thank Xian-Jun Huang and Jing Chen for early involvement of this work. 
This work is supports in part by The National Key R\&D Program of
China No.~2017YFA0402204,
the National Natural Science Foundation of China (NSFC) 
No.~11825506, No.~11821505, No.~U1738209, No.~11851303, No.~11947302, 
% CAS-Xian-Dao
the Key Research Program of the Chinese Academy of Sciences (CAS), Grant NO. XDPB15,
% Shu-Jing project
and the CAS Project for Young Scientists in Basic Research YSBR-006.

\appendix 
\section{Uncertainties in propagation models}\label{app}

\begin{figure}[!h]
\includegraphics[width=0.7\textwidth]{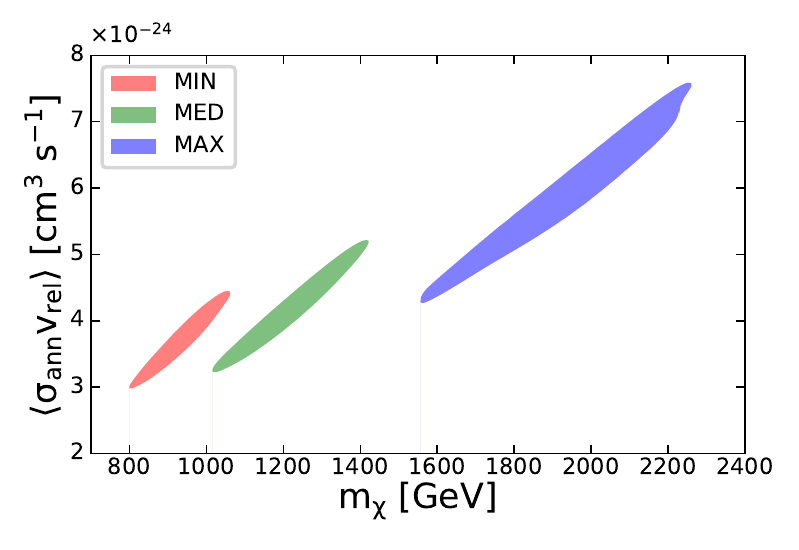}
\caption{Regions favored by the AMS-02 CR positron data in $(m_\chi, \left\langle \sigma_{\rm{ann}} v_{\rm{rel}} \right\rangle)$ plane for $4\mu$ annihilaiton channel at $95\%$ C.L. with $m_\phi=1\ \rm{GeV}$. The red, green, and blue regions correspond to  the ``MIN'', ``MED'', and ``MAX'' propagation models, respectively.}
\label{amspro}
\end{figure}

In this section, we summarize the fitting results of three propagation models ``MIN'', ``MED'', and ``MAX'' mentioned in Sec.~\ref{sec:ams}.
%
%The values of propagation parameters are listed in Tab.~\ref{tab:propagation2}.
%
In Fig.~\ref{amspro}, we show the regions favored by AMS-02 positron data in $(m_\chi, \left\langle \sigma_{\rm{ann}} v_{\rm{rel}} \right\rangle)$ space at $95\%$ C.L. for the three  propagation models.
The posterior means, standard deviations and best-fit values of the fitting parameters and $\chi^2/\text{d.o.f}$ of each propagation model are summarized in Tab.~\ref{tab:promodel}.

\begin{table}[!h]
\begin{tabular}{c|cc|cc|cc|c}
	\hline \hline
	\multirow{3}{*}{Model} & \multicolumn{2}{c|}{$\rm{log_{10}}(m_\chi/\rm{GeV})$} & \multicolumn{2}{c|}{$\rm{log_{10}}(\langle\sigma_{\text{ann}} v_{\text{rel}}\rangle/\rm{cm^3\ s^{-1}})$} & \multicolumn{2}{c|}{$C_{e^+}$} & \multirow{3}{*}{$\chi^2/\text{d.o.f}$}\\
	\cline{2-7} & \multicolumn{2}{c|}{Prior range: [1, 4]} & \multicolumn{2}{c|}{Prior range: [-26, -21]} & \multicolumn{2}{c|}{Prior range: [0.1, 10]}\\
	\cline{2-7} & (Mean, $\sigma$) & Best-fit & (Mean, $\sigma$) & Best-fit & (Mean, $\sigma$) & Best-fit & \\
	\hline
	MIN & $2.97 \pm 0.02$ & 2.97 & $-23.44 \pm 0.03$ & -23.44 & $1.49 \pm 0.02$ & 1.50 & 56.37/32 \\
	MED & $3.08 \pm 0.03$ & 3.08 & $-23.39 \pm 0.04$ & -23.39 & $1.61 \pm 0.03$ & 1.61 & 31.06/32 \\
	MAX & $3.30 \pm 0.05$ & 3.29 & $-23.22 \pm 0.07$ & -23.22 & $1.69 \pm 0.05$ & 1.69 & 22.13/32 \\
	\hline \hline
\end{tabular}
\caption{Prior ranges, posterior means, standard deviations, and best-fit values of DM mass, annihilation cross section, and normalization factor for ``MIN'', ``MED'', and ``MAX'' propagation models with $m_\phi=1\ \rm{GeV}$. The values of $\chi^2/\text{d.o.f}$ are also listed as an estimation of the goodness of fit.}
\label{tab:promodel}
\end{table}

\bibliographystyle{arxivref}
\bibliography{PwaveSommerfeld,inspire_ref,reference,report3}

\end{document}